\documentclass[11pt,a4paper]{article}
\usepackage{jheppub}
\usepackage{amsmath}
\usepackage{hyperref}
\usepackage{amssymb}
\usepackage{mathtools}
\usepackage{mathrsfs}
\usepackage{amsfonts}
\usepackage{bbm}
\usepackage{relsize}
\usepackage{enumerate}
\usepackage{array}
\usepackage{booktabs}
\usepackage{textcomp}
\usepackage{caption}
\usepackage{cite}
\usepackage{cancel}
\usepackage{units}
\usepackage{multirow}
\usepackage{pst-all}
\usepackage{longtable}
\usepackage{relsize}
\usepackage[]{natbib}

\newcommand{\tr}[0]{\ensuremath{\text{tr}}}

\title{ Black Boles  in the 3D
Higher Spin Theory
and Their Quasi Normal Modes}

\author[a,b]{Alejandro Cabo-Bizet}
\author[b]{Edi Gava}
\author[a,b,c]{V. I. Giraldo-Rivera}
\author[c]{K.S. Narain}

\affiliation[a]{SISSA,
Via Bonomea 265, 34128 Trieste, Italy}
\affiliation[b]{INFN, sezione di Trieste, Italy}
\affiliation[c]{ICTP,  Strada Costiera 11, 34014 Trieste, Italy}

\emailAdd{acabo@sissa.it}
\emailAdd{vgiraldo@ictp.it}
\emailAdd{gava@ictp.it}
\emailAdd{narain@ictp.it}

\begin{document}
\title{Black Holes in the 3D 
Higher Spin Theory
and Their Quasi Normal Modes}

\date{\today}
\abstract{
We present a class of 3D black holes based on flat connections which are polynomials in  the 
BTZ  $hs(\lambda)\times hs(\lambda)$-valued  connection. We solve analytically  the fluctuation equations of matter 
in their background and find the spectrum of their Quasi Normal Modes. We analyze  the bulk to boundary two-point functions. 
We also relate our results and those arising  in other backgrounds discussed recently in the literature on the subject.   }

\arxivnumber{}

\maketitle

\flushbottom
\section{Introduction}
A distinctive feature of  black holes,  in both, asymptotically flat and asymptotically (A)dS space-times,
is the existence of quasi normal modes:  if one perturbs a black hole, one finds damped modes,
i.e. modes whose frequencies are complex,  signalling the fact that the corresponding field can decay by 
falling  into the black hole.   In the AdS case these modes have an interpretation in the dual CFT 
as describing the approach to equilibrium of the perturbed thermal state \citep{Horowitz:1999jd, Cardoso:2001bb, Son:2002sd}. This phenomenon has been
studied extensively,  especially after the proposal of the AdS/CFT correspondence, in the ordinary (super)gravity
context in various dimensions. In particular, for gravity  coupled to various matter in $D=3$,  the case of the BTZ black hole has been studied in detail.  

We will be interested in generalising  the problem to the context of higher spin systems in $D=3$. Such
systems, with finite number of higher spins $\leq N$,  can be formulated via Chern-Simons theories based on $sl(N)$ algebras,
but, like ordinary 3D gravity, they do not contain propagating degrees of freedom and, moreover,  they do not allow coupling to propagating matter. In order
to introduce (scalar)  matter coupled to the higher spin sector,  one  formulates the theory in terms of a flat connection $(\mathcal{A},\mathcal{\bar{A}})$
for the infinite dimensional algebra $hs(\lambda)\times hs(\lambda)$  \citep{Fradkin:1987ks,Prokushkin:1998bq,Vasiliev:1999ba}. The matter fields are packaged in an algebra valued 
master field $C$,  a section  obeying a horizontality condition with respect to the flat connection, in a way that will be detailed below.
It turns out that if one embeds the BTZ black hole in this system, one can follow a ``folding" procedure to reduce the equation of motion for $C$ in the BTZ
background to an ordinary second order equation for the lowest, scalar,  component of the field $C$, with a $\lambda$ dependent mass, $m^2=\lambda^2-1$. Therefore
the corresponding quasi normal modes are the usual ones that one finds for a massive scalar field coupled to BTZ in the ordinary gravity case.

However, the higher spin systems are expected to admit generalised black holes carrying different charges, other than the mass and angular momentum
carried in the  BTZ case \citep{Gutperle:2011kf, Ammon:2011ua,Kraus:2012uf, Beccaria:2013yca, Banados:2012ue, David:2012iu,Ferlaino:2013vga, Perez:2014pya, Perez:2012cf, Compere:2013gja, Gutperle:2013oxa, deBoer:2013gz}. 
The issue then arises to study matter fluctuations in their background and possibly identify the spectrum of the corresponding quasi normal modes. Unlike for the BTZ background, one expects in general
the ``folding" procedure to give rise to a differential equation of order higher than two for the matter scalar field  and it is not  to be expected to be able to solve it
analytically.

In this paper we present a class of flat connections, 
depending on parameters ($\mu, \bar{\mu}$), in such a way that when  $\mu$ and  $\bar{\mu}$ go
to zero we recover the BTZ connection.
We will argue that they correspond
to black hole configurations in 3D $hs(\lambda)\times hs(\lambda)$
higher spin gravity. In addition, we will be able to solve the equations for  matter fluctuations 
analytically and therefore identify the presence of quasi normal modes. 

As first discussed in  \citep{Gutperle:2011kf}, establishing whether a given geometry represents a black hole in higher spin theories
is a subtle issue, due to the presence of a higher spin gauge degeneracy 
that can, to mention an example, 
relate seemingly black hole geometries to geometries without horizons. We will follow the criterion of \citep{Gutperle:2011kf}
and impose the BTZ holonomy conditions on the connection around  the euclidean time $S^1$.
As a result spacetime tensor fields \citep{Campoleoni:2010zq} will be  shown to behave smoothly at the  horizon.

But as remarked above, further evidence arises from the analysis of their interaction with matter, in particular from the existence of  quasi normal modes and their dispersion relations.
Another subtle and important issue, whose general aspects  have been subject of recent investigations, with different conclusions, \citep {Bunster:2014mua, Perez:2014pya, Perez:2012cf, Compere:2013gja, deBoer:2014fra},  concerns the precise determination of the charges carried by our backgrounds and, more generally, their asymptotic symmetry algebras. Perhaps,  one could get a clue of the general answer by studying the truncations of $hs(\lambda)$ with integer $\lambda$, we hope to come back to this problem in the future. In this way one would be able of, first, properly define their charges and, second, identify whether they are of higher spin character or not.

As for the bulk to boundary 2-point function, even though the differential equations of motion that determine them are of order higher than two, they are described by combinations of pairs of solutions of a 2nd order PDE's. Only one of all these pairs is smoothly related to the solutions corresponding to
a real scalar field with $m^2=\lambda^2-1$ propagating in the BTZ black hole \citep{Didenko:2006zd,Prokushkin:1998bq}, as $\mu,\bar{\mu}\rightarrow 0$. 

The outline of the paper is as follows. First, in section \ref{sec:BH} we introduce the ans\"atze mentioned above and show that  they define smooth horizons by use of the relation connections - metric-like fields proposed in \citep{Campoleoni:2010zq}. Then, we  identify our ($\mu$,$\bar{\mu}$) with the so called chemical potentials in the solutions introduced in \citep{Gutperle:2011kf} and generalisations to $hs(\lambda)$ of the $sl(3,\mathbb{R})$ black holes with higher spin charges (and $W_3\times W_3$ asymptotic symmetry algebra) presented by the authors in \citep{Bunster:2014mua}, that from now on, we denote as GK and BHPT2\footnote{These, are solutions in $hs(\lambda)\times hs(\lambda)$ with chemical potentials turned on in the time direction. In this way the fixed time asymptotic symmetry algebra is precisely $W_\lambda \times W_\lambda$. Details on their construction are given in section \ref{sec:BH}.}, respectively. We do it  by identifying the gauge transformation relating our connection to  those ans\"atze. Next, in section \ref{sec:ex} we discuss the  equations of motion for the effective scalar in the black holes
and  describe the strategy  to solve them for a generic element in the class. We  give the explicit solutions  for a couple of particular cases. An important fact to stress on, is that even though connections and generic metric like fields do break the asymptotic of $AdS$, the equations for fluctuations do preserve the behaviour of minimally coupled scalars.
 
In section \ref{sec:QNM} we show how to obtain the quasi normal modes and bulk to boundary 2-point functions for a generic element in the class and discuss them in the same  particular cases. As a last check, we transform our results to the GK and 
BHPT2 descriptions and verify that the result of the gauge transformation coincides with the perturbative solution of the equations of motion for linear fluctuations of matter, written in those ans\"atze, 
as it should. \\

\section{A class of black holes in the  $hs(\lambda)\times hs(\lambda)$  theory}\label{sec:BH}

In this section of the paper we argue that a given class of $hs(\lambda)\times hs(\lambda)$ flat connections do have a space time interpretation as black holes. As a first argument, we resort to the usual relation between connections and metric like tensor fields discussed in the finite dimensional case  in \citep{Campoleoni:2010zq}. 
 
We start by writing down the generic form for the flat connections of interest:
\begin{eqnarray}
\mathcal{A}_\rho=V^{2}_0, ~~ \bar{\mathcal{A}}_\rho=-V^{2}_0, ~~~~~~~~
\nonumber\\
\mathcal{A}_{t,\phi}= b A_{t,  \phi} b^{-1},  ~~   \bar{\mathcal{A}}_{t,\phi}=\bar{b} \bar{A}_{t,\phi} \bar{b}^{-1}, \label{Arho}
\end{eqnarray}
with $b=e^{- \rho V^2_0}$, $\bar{b}=e^{\rho V^2_0}$. The generators and structure constants for $hs(\lambda)$ algebra are listed in appendix \ref{app:Conv}.
Let us denote our space-time coordinates as $(\rho,t,\phi)$  and  restrict our analysis to connections that obey the gauge choice \eqref{Arho} with $A$ independent of $x_a=(t,\phi)$. 

The relation between the  connection and the space time tensor fields is:
\begin{equation}
 g^{(n)}=-\frac{1}{2} tr(e^n), ~~~~ e=\mathcal{A}-\bar{\mathcal{A}},  \label{tensor}
\end{equation}
with $e$ being the  dreibein. As a starting point we remind  the condition:
\begin{equation}\label{et} e_t |_{\rho=0}=0,\end{equation} required in order to have a smooth horizon at $\rho=0$ in the spacetime tensor field  $g^{(n)}$. Under (\ref{et}) each $t$ component in $g^{(n)}$ will  have  a  zero at $\rho=0$ with the appropriate order. By appropriate orders we mean those that make the corresponding reparameterisation
invariant quantities smooth at $\rho=0$.  For instance, $g_t^{(n)}\sim \rho^n$, and thence, it will be smooth after transforming to a regular coordinate system about the horizon.  In virtue of (\ref{Arho}) we can rewrite (\ref{et}) as:
\begin{equation}\label{et0}
\bar{A}_t =A_t.
\end{equation}
From the  flatness condition the $\phi$ components are constrained to be of the form:
\begin{equation}
A_{\phi}=P\left(A_t\right), \,  \bar{A}_{\phi}=\bar{P}\left(A_t \right) \footnote{ It could be the case that $A_t=P(A_\phi)$ and not the other way around, but for our purposes we stick to the case written above. In fact the most general case is $A_{\phi}=P_\phi(A)$ and $A_{t}=P_t(A)$ with a generic $A\in hs(\lambda)$.},\end{equation}\\
where we take $P$ and $\bar P$  to be polynomials in $A_t$ and $\bar{
  A}_t$ respectively.
The condition:
\begin{equation}\label{even}
g^{(n)}(\rho)=g^{(n)}(-\rho),
\end{equation}
guarantees that all the  components of  $g^{(n)}$ will be smooth in the Cartesian coordinates in the plane $(\rho,t)$, with $\rho$ representing the radial coordinate. Condition (\ref{even}) ensures smoothness for the $g^{(n)}$ at $\rho=0$. As far as Euclidean conical singularity is  concerned, it will be automatically excluded by requiring fulfilment of the BTZ holonomy condition \citep{Gutperle:2011kf}. See the paragraph before (\ref{conical}) for more details.

Let us identify a sufficient condition on the connections ($A, \bar{A}$) for (\ref{even}) to hold. Consider the generic connections:
\begin{eqnarray}
A_{a}=\sum_{(s,m_s)}c^{s}_{m_s} V^{s}_{m_s}, \, \bar{A}_{a}=\sum_{(s,m_s)}\bar{c}^{s}_{m_s} V^{s}_{m_s}. \label{condMod}
\end{eqnarray}
Notice that the change $\rho$ to $-\rho$ in (\ref{condMod}) is equivalent to the change $V^{^s}_{m_s}\rightarrow V^{^s}_{-m_s}$\footnote{Here we consider $s=1,\ldots \infty$, $m_s=-2s+1,\ldots,2s-1$. So that under summation the indices $s$ and $m_s$ are mute and can be renamed without lack of rigor.}. 

By inserting (\ref{condMod}) in (\ref{tensor}), and using the properties of
the $\star$-product, we can notice that  $tr( e_a ^n)$  is invariant under
the combined action of $\rho \rightarrow -\rho$ and any of the
following pair of $\mathbb{Z}_2$ transformations:
\begin{eqnarray}\label{Z2sym}
\text{I}: c^{s}_{m_s} \left(\bar{c}^{s}_{m_s}\right)\rightarrow c^{s}_{-m_s}\left( \bar{c}^{s}_{-m_s} \right) & \text{ AND/OR  }~~ ~ \text{I} \times \text{II}, \end{eqnarray}
with the $\mathbb{Z}_2$ II given by
\begin{equation}
\text{ II }: \bar{c}^{s}_{-m_s}\rightarrow -c^{s}_{m_s}.
\end{equation}

Transformation I together with $V^s_{m_s}\rightarrow V^s_{-m_s}$ leaves
the dreibein $e_a=\mathcal{A}_a-\bar{\mathcal{A}}_a$ invariant and
therefore  the trace of  powers of $e_a$. The transformation II leaves  $tr (e_a^n)$ invariant but generically not  the dreibein $e_a$. 

A trivial (even) representation of $(A, \bar{A})_a$ under
\eqref{Z2sym} is sufficient condition for \eqref{even}. Should some
components in $(A,\bar{A})$ not remain invariant under the
$\mathbb{Z}_2$  I or I$\times$II,  but carry a non trivial (odd)
representation under any of them, then the corresponding component of
the dreibein $e$  will carry a non trivial (odd) representation
too. Condition \eqref{even} will thus imply that traces involving an odd number of such  components  must vanish. 

Let us analyze the particular case of the BTZ connection
\begin{eqnarray}\label{btz}
A_t=\bar{A}_t=\frac{1}{2}a, \  \  A_\phi=-\bar{A}_\phi=\frac{1}{2}a,
\end{eqnarray}
where
\begin{equation}
a=V^{2}_1+M V^{2}_{-1}.
\end{equation}
 From now on, for simplicity, we will choose the value $M=-1$, which locates the horizon at $\rho=0$. For later use we define $a_{\pm \rho}=b a b^{-1}$.
 
The $\phi$ component of the pair $(A,\bar{A})$ remains invariant under the transformation II whereas the $t$ component is odd. However the $t$ component is also odd under I and so even under the composition I$ \times$ II. Finally, the following symmetries of the corresponding $t$ and $\phi$ components of the dreibeins
 \begin{eqnarray}
 e_t&=& \frac{1}{2}(a_{\rho}-a_{-\rho})\equiv a_{\text{ I }\times\text{ II } -even},\nonumber \\
 e_\phi&=&\frac{1}{2}(a_{\rho}+a_{-\rho})\equiv a_{\text{ II}-even},
  \end{eqnarray}
imply that \eqref{even}  holds for the connection
\eqref{btz}. We can still get further information from symmetries. As
$e_t$ and $e_\phi$ are odd under I, any tensor field component
with an odd number of $t$ plus $\phi$ directions, vanishes.  As $e_t$
and $e_\phi$ are odd and even respectively, under II, any tensor component with an odd number of $t$ components vanish. Finally, what said before implies that any tensor component with and odd number of $\phi$ directions vanish too. 

Much of what was used for the BTZ case before, holds also for generic connections. Specifically:
\begin{itemize}
\item Any pair of connections ($A,\bar{A}$) that carries a trivial representation under I or I $\times$ II, will define metric-like fields obeying \eqref{even}. 
\end{itemize}

Additionally, one can argue also for a necessary
condition for (\ref{even}) to hold. Let us suppose that a pair $(A,\bar{A})$ contains a part
$(A^{rep},\bar{A}^{rep})$ that satisfies the conditions above, and a part $(\delta
A,\delta\bar{A})$ that does not, but
still defines metric like fields which are even under $\rho$ to
$-\rho$. In that case the term $(\delta \mathcal{A}-\delta
\bar{\mathcal{A}})$ should be orthogonal to itself\footnote{ The
  orthogonality is meant with respect to the trace operation in $hs(\lambda)$.}, its
powers, and powers of the generators in
$(\mathcal{A}^{rep}-\bar{\mathcal{A}}^{rep})$ (This is possible
to find,  for example $V^3_2$ is orthogonal with itself and its
powers). Should this not be the case, the term ($\delta \mathcal{A}-\delta \bar{\mathcal{A}}$) would give contributions which are not even in $\rho$ (based on the invariance property of the trace mentioned above). However, if $(\mathcal{A}^{rep}-\bar{\mathcal{A}}^{rep})$ contains all of the $sl(2,\mathbb{R})$ elements, $V^2_{0,\pm1}$, it is impossible to find  a set of generators in $hs(\lambda)$ that is orthogonal to every power of them. In that case, symmetry under any of the $\mathbb{Z}_2$ transformations in the maximal set, out of the \eqref{Z2sym}, $(\text{I},\text{ I}\times \text{ II})$ for any $(s,m_s)$\footnote{Notice that there are many possible $\mathbb{Z}_2$'s. The number grows exponentially with the number of generators in ($\mathcal{A}-\bar{\mathcal{A}}$). The calligraphic letters indicate the full connection, $\rho$ component and $\rho$ dependence included.} is also a necessary condition for \eqref{even}. 


At this point we specify our class of connections:
\begin{eqnarray}
A_t=\bar{A}_t =P_t\left(a\right), ~~~~~~~~\nonumber \\
A_\phi= \frac{1}{2}a+P_\phi\left(a\right), ~~  \bar{A}_\phi=-\frac{1}{2}a+\bar{P}_\phi\left(a\right), \label{BHoles}
\end{eqnarray}
with $P_t$, $P_\phi$ and $\bar{P}_\phi$ being arbitrary traceless polynomials of the form
\begin{eqnarray}
P_t=\sum^{\infty}_{i=0}\nu_i \left(a^{2i+1}-\text{ trace }\right),~~~~~~~~~~~~~~~~~~~\nonumber\\
P_{\phi}=\sum^{\infty}_{i=0}\mu_{i+3} \left( a^{2i+2}-\text{ trace }\right), ~~ \bar{P}_{\phi}=\sum^{\infty}_{i=0}\bar{\mu}_{i+3} \left( a^{2i+2}-\text{ trace }\right).\label{backgrounds}
\end{eqnarray}
Notice that (\ref{backgrounds}) obeys (\ref{et}) and that $P_t$ and $P_\phi$ are selected in such a way that $g_{t\phi}=0$. We also choose the components $g_{\rho t }$ and $g_{\rho\phi}$ to vanish. In particular (\ref{backgrounds}) reduce to the non rotating BTZ$_{M=-1}$ connection in the limit $\nu_0=\frac{1}{2}$, $\nu_{i>0}=0$,  and vanishing $\mu_i$, $\bar{\mu}_i$. Now:
\begin{itemize}

\item The transformations of $a$, the corresponding deformation polynomials
$(P_{\phi}(a)$, $\bar{P}_{\phi}(a)$, $P_t(a))$ and the $\rho$ components $\pm V^2_0$ under  I  in \eqref{Z2sym}, are odd, even, odd and even respectively. 
\item In virtue of properties of the $\star$-product, the traces with odd numbers of $a$ and $P_t(a)$ with any number of insertions of $V^2_0$ and $(P_{\phi}(a),\bar{P}_{\phi}(a))$,  vanish, and so all non vanishing traces are even under I and henceforth even under $\rho \rightarrow-\rho$.
 \end{itemize}
 We conclude that the ans\"atze (\ref{backgrounds}) give rise to spacetime tensor fields that obey  (\ref{even}). In fact we explicitly checked \eqref{even} to hold up to arbitrary higher order in $n$ and the order of the polynomials $P$ and $\bar{P}$.

In the near horizon expansion, $g^{(2)}$, the line element defined by (\ref{tensor}), will look like:
\begin{equation}d\rho^2-\frac{4}{T^2} \rho^2 dt^2+\ldots =\rho^*dv^2+\frac{1}{2}d\rho^* dv +\ldots,\label{lineElement}
\end{equation}
with $v=t-\frac{ T}{2} log(\rho)+\ldots$ and $\rho^*=\frac{4}{T^2} \rho^2+\ldots$ being coordinate redefinitions that are going to be useful later on when analyzing fluctuations. The $\ldots$ denoting higher orders 
corrections in $\rho$. The temperature:
\begin{equation}\label{temperature}
T\left(P_t\right)\equiv \frac{1}{\sqrt{\frac{1}{2}tr\left( [ P_t(a),V^2_0 ]^2\right)}},\footnote{From the positiveness of the traces $tr(V^{2s}_{2m_s+1}V^{2s}_{-2m_s-1})$, see (\ref{trace}), in the interval $0<\lambda<1$ and the fact we have chosen odd powers of $a$ in $P_t$ it follows that the quantity inside the roots in (\ref{smoothness2}) and (\ref{smoothness3}) is a sum of positive defined quantities and hence positive defined. We stress that we restrict our study to the interval $0<\lambda<1$.}
\end{equation}
defines the thermal periodicity under $t\rightarrow t+ \pi T i$.

We will focus our study in the cases $\nu_0=\frac{1}{2}$, $\nu_{i>0}=0$. These are solutions that obey the usual BTZ holonomy-smoothness condition as the temporal component of the connection coincides with the BTZ one with $M=-1$. This implies that not only the eigenvalues of the time component of connection  are  the same as BTZ$_{M=-1}$, but also that  the holonomy around the contractible euclidean time cycle coincides
with the BTZ case, since the euclidean periodicity, determined by the temperature $T\left(\frac{1}{2} a\right)=2$, is the same as for the BTZ$_{M=-1}$.  
 
 However before going on, let us comment on the possibility of arbitrary $\nu_i$.  The euclidean smoothness condition is:
\begin{equation}
e^{\pi i T(P_{t}) P_t(a)}\sim V^1_0. \label{smoothness2}
\end{equation} 
To solve for (\ref{smoothness2}) we use the fact that $\pi i P(a)$, with $P(a)$ an arbitrary polynomial in $a$ with arbitrary integer coefficients, are known to exponentiate to $V^1_0$ in the region $0< \lambda<1$, see \citep{Monnier:2014tfa}. 

Then relations \eqref{smoothness2} reduce to find out the $\nu_i$ such that $\nu_i T(P_t)$ are integers.
To study this quantization conditions it is useful to write down $P_t$ in the basis
\begin{eqnarray}
a_{\perp}^{s-1}&\equiv&\frac{1}{N_s} \sum_{t=0}^{s-1}(-1)^t
\left(\begin{array}{cc}  
s-1 \\t\end{array}\right) V^{s}_{s-1-2t} ~\sim  (a^{s-1})\big|_{V^{t<s}_{m_t}\rightarrow 0},\label{perpbasis}
\end{eqnarray}
where $N_s$ is a normalization factor, chosen in such a way that:
$\tr((a_{\perp}^{s-1})^2)=1$. We get thus
\begin{eqnarray}
P_t(a)&=&\sum_{s=0}^{\infty}{\nu_{\perp}}_s \frac{a^{2s+1}_{\perp}}{\sqrt{\frac{1}{2}tr([a^{s-1}_{\perp},V^2_0]^2)}}, ~~~~ \nu^s_{\perp}
=M^{si}\nu_i,
\label{smoothness3}
\end{eqnarray}
where the linear transformation matrix $M$  is upper triangular. In the appendix \ref{sec:M} we present the explicit form for $M$, (\ref{M}), for the case $\mu_{2i+1}\neq0$, with $i=0,\ldots,4$.
An important property to use is that the eigenvalues (the diagonal elements) of $M$ can be checked to be larger or equal than 1 in the range $0<\lambda<1$ until arbitrary large $i$.  

The desired quantization conditions can be written  as:
\begin{equation}
\nu_i T(P_t)= ({M^{-1}})_{is} \cos{\theta^s}=n_i, \label{qn}
\end{equation}
with $\cos{\theta^s}\equiv \frac{\nu_\perp^s}{\sqrt{\sum_s \left(\nu^s_{\perp}\right)^2}}$ and $n_i$ an arbitrary integer. The condition for the quantization relation \eqref{qn} to admit solutions is:
 \begin{equation}
\sum_{s=1}^\infty {(M\centerdot n)^s}^2=1. \label{quantis}
\end{equation}
In appendix \ref{sec:M} we show that the property of the eigenvalue of $M$ mentioned above excludes  the presence of other solutions to the consistency condition (\ref{quantis}) in the region $0<\lambda<1$, apart from the trivial one, $n_0=1$ ($\nu_0=\frac{1}{2}, ~\nu_{i>0}=0$). Here, we just continue with the cases that are continuously linked to the BTZ connection in the limit $\mu_i$, $\bar{\mu}_i$ to zero. Namely $\nu_0=\frac{1}{2},~ \nu_{i>0}=0$.  The requirement of the BTZ holonomy condition will guarantee the absence of any possible conical singularity in the tensor like fields as the dreibein itself is thermal periodic.


Generically, \eqref{backgrounds} will define asymptotically Lifshitz metrics with critical exponent $z<1$, except for the cases in which the contributions out of the deformation parameters $\mu_i$, $\bar{\mu}_i$ will not provide $\rho$ dependence. An example being when 
$\bar{\mu}_i=0$ (or $\mu_i=0$)
in which case the only contribution to $g_{\phi\phi}$ comes at quadratic order in $\mu_{i}$(or $\bar{\mu}_{i}$) but it is independent of $\rho$ due to the cyclic property of the trace. In those cases the metric becomes asymptotically AdS. 

To summarize, \eqref{backgrounds} will define metrics of two classes:
\begin{itemize}
\item Generically Lifshitz metric with $z<1$.
\item AdS metrics when $\mu_{2i}=0$ (or $\bar{\mu}_{2i}=0$). 
\end{itemize}

This classification relies on the definition (\ref{tensor}). 
For instance the line elements coming from (\ref{tensor}) for the cases $\mu_3\neq 0$,  $\bar{\mu}_3=-\mu_3\neq0$ and $\bar{\mu}_3=\mu_3\neq0$
look like :\\
\begin{eqnarray}
ds_{(\mu_3,~0)}^2&=&d\rho^2-\sinh^2  \rho~ dt^2+\left( \cosh^2 \rho+\frac{16(\lambda^2-4)}{15}  \mu_3^2\right)d\phi^2,\nonumber \\
ds_{(\mu_3,-\mu^3)}^2&=& d\rho^2 - \sinh^2 \rho~ dt^2+\frac{1}{30} \left(12 \left(\lambda ^2-4\right) \mu_3^2 \cosh (4 \rho )\right. \nonumber\\&&\left.~~~~~~~~~~~~~~~~~~~~+5 \left(4 \left(\lambda ^2-4\right)\mu_3^2+3 \cosh (2 \rho
   )+3\right)\right) d \phi^2, \nonumber  \\
   ds_{(\mu_3,~\mu_3)}^2&=& d\rho^2 - \sinh^2 \rho ~dt^2+\frac{1}{5} \cosh ^2(\rho ) \left(-8 \left(\lambda ^2-4\right) \mu _3^2 \cosh (2 \rho )\right. \nonumber \\ &&\left.~~~~~~~~~~~~~~~~~~~~~~~~~~~~~~~~~~~~~~~~~~+8 \left(\lambda ^2-4\right) \mu _3^2+5\right)d \phi^2.\nonumber\\
\label{eq:lifini}
\end{eqnarray}
 The first line element in (\ref{eq:lifini}) behaves asymptotically as
 AdS$_3$ and shows a smooth horizon at $\rho=0$, while the last two
 cases are Lifshitz metrics with dynamical critical exponent  $z=\frac{1}{2}<1$. Should we
 have turned on a higher spin $\mu$ deformation, the parameter $z$
 would have decreased like  $z=\frac{1}{4},\frac{1}{8}\dots$.  
 
 The bulk of the present study, section \ref{sec:ex}, will be devoted to the study of matter fluctuations
 around the connections (\ref{BHoles}), which  are not just gravitational but involve also higher spin tensor fields turned on. This further analysis will confirm the expectation that these backgrounds truly describe black holes, through the ``dissipative" nature of matter fluctuations we will find.

Before closing this section, we  make contact (perturbatively in $\mu_3$) with other relevant backgrounds studied in the literature recently. More precisely, we look  for static gauge parameters ($\Lambda, \bar{\Lambda}$) (independent of $x_{1,2}$), that transform \eqref{backgrounds} to the GK \citep{Gutperle:2011kf} and BHPT2 backgrounds mentioned in the introduction (those that generalise the $sl(3,\mathbb{R})$ black holes with higher spin charge introduced in \citep{Bunster:2014mua, Henneaux:2013dra} to $hs(\lambda)$). Notice that these gauge transformations will not change the eigenvalues of the components ($A_{1,2},\bar{A}_{\bar{1},\bar{2}}$) of the connections because they are just similarity transformations. The two classes of backgrounds we want to relate ours to, are described by
the following connections:  
 \begin{eqnarray}
 A_1&=&V^2_1+\mathcal{L}V^2_{-1}+\mathcal{W}V^3_{-2}+\mathcal{Z} V^4_{-3}+\ldots, ~~A_2=\sum^\infty_{i=0}\mu_{i+3}\left(A_1^{i+2}-\text{ traces} \right), \nonumber\\ 
\bar{A}_{\bar{1}}&=&V^2_{-1}+\bar{\mathcal{L}}V^2_{1}+\bar{\mathcal{W}}V^3_{2}+\bar{\mathcal{Z}}V^{4}_3+\ldots,~~~   \bar{A}_{\bar{2}}=\sum^{\infty}_{i=0}\bar{\mu}_{i+3}\left(\bar{A}_{\bar{1}}^{i+2}-\text{ traces }\right). \nonumber\\\label{KGPBT}
 \end{eqnarray}
Our parameters $(\mu_i,\bar{\mu}_i)$ will be identified precisely with the chemical potentials in \eqref{KGPBT}. In our approach the charge-chemical potential relations \citep{Gutperle:2011kf,Ammon:2012wc} are determined a priori by the condition $\nu_0=\frac{1}{2}, ~ \nu_{i>0}=0$. Namely, after applying the gauge transformations $(\Lambda,\bar{\Lambda})$ the charges $\mathcal{L}$, $\mathcal{W}$ and $\mathcal{Z}$ will be already written in terms of the chemical potentials $({\mu}_i,\bar{\mu}_i)$. In this way one can generate GK, and BHPT2  ans\"atze with more than one $(\mu_i,\bar{\mu}_i)$ turned on, and with the holonomy conditions already satisfied. However, with the choice $\nu_0=\frac{1}{2}$, $\nu_{i>0}=0$ one can only reach branches that are smoothly related to the BTZ$_{M=-1}$.\\
Taking $x_1=x_{\bar{2}}=x_+$ and $x_2=x_{\bar{1}}=x_-$,  we recover the GK background, whereas for  $x_1=x_{\bar{1}}=\phi$ and $x_2=x_{\bar{2}}=t$ we get BHPT2.

For later use, we write down the particular gauge transformations that takes the representative with non vanishing $\mu_3=-\bar{\mu}_3$ into the wormhole ansatz  for GK's case. They read,
respectively, to leading order in $\mu_3=-\bar\mu_3$:
 \begin{eqnarray}
 \Lambda_{GK}&=&\mu_3\left(-\frac{5}{3}e^{-\rho} V^3_{-1}+e^\rho V^3_1\right)+~{\rm commutant }~{\rm of} ~ a_{\rho}+ O(\mu_3^2),\nonumber \\
 \bar{\Lambda}_{GK}&=&\mu_3 \left(e^\rho V^3_{-1}-\frac{5}{3}e^{-\rho}V^3_1\right)+~{\rm  commutant}~{\rm of} ~ a_{-\rho}+ O(\mu_3^2).\label{LKGP}
 \end{eqnarray}
 The holonomy conditions are satisfied a priori and so the corresponding charge-chemical potential relations are as follows:
 \begin{eqnarray}
 \mathcal{L}=\bar{\mathcal{L}}=-1+O(\mu_3^2), ~ \mathcal{W}=-\bar{\mathcal{W}}=\frac{8}{3}\mu_3+O(\mu_3^3), ~ \mathcal{Z}=\bar{\mathcal{Z}}=O(\mu_3^2), ~\ldots \label{KGP}
 \end{eqnarray}
For BHPT2, namely when the chemical potentials are turned on along the $t$ direction and the asymptotic symmetry algebra is the undeformed $W_\lambda\times W_\lambda$ \citep{Henneaux:2013dra,Gary:2014mca}, they are given by: 
\begin{eqnarray}
\Lambda_{BHPT2}&=&2\Lambda_{GK}+ O(\mu_3^2)\nonumber, \\
\bar{\Lambda}_{BHPT2}&=&2\bar{\Lambda}_{GK}+ O(\mu_3^2).\label{LBT}
\end{eqnarray}
In this case the relations charge-chemical potential are:
\begin{equation}
 \mathcal{L}=\bar{\mathcal{L}}=-1+O(\mu_3^2), ~ \mathcal{W}=-\bar{\mathcal{W}}=\frac{16}{3}\mu_3+O(\mu_3^3), ~ \mathcal{Z}=\bar{\mathcal{Z}}=O(\mu_3^2) . \label{BT}
\end{equation}
Later on, we will apply these transformations to the matter fluctuations in the  $\bar{\mu_3}=-\mu_3\neq0$ background in (\ref{backgrounds}). 

\section{Equations for fluctuations } 
\label{sec:ex}
In this subsection we show how to obtain the differential equations  for the scalar fluctuations  over the backgrounds (\ref{backgrounds}). 
Firstly, we review how this works for the BTZ$_{M=-1}$ case. This will allow us to identify a strategy for the cases (\ref{backgrounds}). 

As mentioned in the introduction, the equation of motion 
of the master field $C$ in generic background connections $(A,\bar{A})$ is simply the horizontality condition:
\begin{equation}\label{EqFluc}
\tilde{\nabla} C\equiv dC+\mathcal{A} \star C-C\star
\overline{\mathcal{A}}=0 \,\, \text{ with }  \,\, C=\sum C^{s}_{m_s}V^{s}_{m_s},
\end{equation}
whose formal solution and its corresponding transformation law under left multiplication $(g,\bar{g})\rightarrow(e^\Lambda g, e^{\bar{\Lambda}}\bar{g})$, are,
respectively:
\begin{equation}\label{sol}
C= g \, \mathcal{C} \,\overline{g}^{-1}  \text{    and} ~ C_{(\Lambda,\bar{\Lambda})}=e^\Lambda C e^{-\bar{\Lambda}},
\end{equation}
where $d\mathcal{C}=0$ and $\mathcal{C}=\sum
\mathcal{C}^{s}_m V^{s}_m$. 

The trace part of the master field $C$ and its transformation law are also:
\begin{equation}
C^{1}_0=(C)\big|_{V^1_0}\, \text{ and } ~ {C^1_0}_{(\Lambda,\bar{\Lambda})}=\left(e^{(\Lambda-\bar{\Lambda})}C\right)\big|_{V^1_0}.\label{C10}
\end{equation}
The integration constant $\mathcal{C}$ is evaluated in the limit $C\big |_{g\rightarrow 1}$. In our cases (\ref{backgrounds}) $g$ goes to $1$ at the points $(\rho,x_a)=0$. However notice that these points are located at the horizon $\rho=0$ of (\ref{backgrounds}) and, as we shall see, many of the components of the master field $C$ will diverge there.
 
Our aim  is to ``fold" \eqref{EqFluc} for our ans\"atze (\ref{backgrounds}) with $\nu_{0}=\frac{1}{2}$, $\nu_{i>0}=0$. By ``folding" we mean the  process of expressing
 every $C^s_{m_s}$ in terms of $C^1_0$ and its derivatives, and finally to obtain  a differential equation for $C^1_0$. 
For such a  purpose  we start  by reviewing how this process works for the simplest case, BTZ$_{M=-1}$,  and 
in doing so we will discover how  to fold the matter fluctuations in the case of the backgrounds (\ref{backgrounds}). 

We start by proving that for BTZ$_{M=-1}$ every higher spin component $
 C^{s}_{m_s}$, can be expressed in terms of  $\partial_{\pm}$ derivatives of $C^1_0$ and $C^2_0$. Using the explicit forms for $g$ and $\bar{g}$  in this case:
 \begin{equation}\label{masterfbtz}
 C=e^{-a_{\rho}x_+}  \mathcal{C}(\rho) e^{-a_{-\rho} x_-}.
 \end{equation}
 It is easy to see that:
 \begin{eqnarray}
 \partial_{\pm} C^1_0=-( a_{\pm \rho}C)\big|_{V^1_0}\sim -(e^{\pm \rho} C^{2}_{1}- e^{\mp \rho} C^{2}_{-1}),\label{dplusphi}
 \end{eqnarray}
 from where \eqref{C21} of the Appendix \ref{sec:HOEqs} is immediate. By $\big(\ldots\big)\big|_{V^1_0}$ we denote the coefficient of $V^1_0$ in $\big(\ldots\big)$. 

 Now we can repeat the procedure at second order in $\pm$ derivatives
 of $C^1_0$. At this stage we can write down three
 combinations:  \[\partial^2_{+}, \ \partial^2_{-}, \ \partial^2_{+-}, \] which would generate the following quadratic relations inside the trace element:
 \begin{eqnarray} \label{array1}
 a_{\rho}^2&=& \tilde{V}^{1}_0+e^{2\rho} V^{3}_2-2V^{3}_0+e^{-2
   \rho}V^{3}_{-2},\\ \label{array2}
 a_{-\rho}^2&=& \tilde{V}^{1}_0+e^{-2\rho} V^{3}_2-2V^{3}_0+e^{2
   \rho}V^{3}_{-2}, \\
 a_{\rho} a_{-\rho}&=& \cosh{2\rho}(\tilde{V}^{1}_0-2V^{3}_0)-2 \sinh{2\rho}V^{2}_0+V^{3}_2+V^{3}_{-2}, \label{array3}
 \end{eqnarray}
where $\tilde{V}^{1}_0=\frac{\left(\lambda^2-1\right)}{3}V^{1}_0$. 

Equations \eqref{array1}, \eqref{array2} and \eqref{array3}, allow to
write down $C^{3}_{-2}$, $C^{3}_{0}$ and $C^{3}_{2}$ in terms
of  \[ \left( \partial^2_{+} C^1_0,\ \partial^2_{-}C^1_0,
  \ \partial^2_{+-} C^1_0,  C^{2}_{0}\right), \]
so that one arrives to the relations \eqref{C30} and \eqref{C32}.

Proceeding this way,  we see that  at the level $s=3$ we can still use first derivatives acting on $C^{2}_0$:
\begin{eqnarray}\label{dplusphi2}
\partial_{+} C^{2}_0=-(V^{2}_0 a_{\rho} C)\big|_{V^1_0} \,\, \text{  and  } \, \, \partial_{-} C^{2}_0=- ( a_{-\rho} V^{2}_0 C)\big|_{V^1_0}.\label{dminusphi2}
\end{eqnarray}
Then, if we use:
\begin{eqnarray}
V^{2}_0 a_{\rho}&=&-\frac{1}{2}(e^{\rho}V^{2}_{1}+e^{-\rho}V^{2}_{-1})-e^{-\rho}V^{3}_{-1}+e^{\rho}V^{3}_{1},\\
a_{-\rho} V^{2}_0 &=&\frac{1}{2}\left(e^{-\rho}V^{2}_{1}+e^{\rho}V^{2}_{-1}\right)-e^{\rho}V^{3}_{-1}+e^{-\rho}V^{3}_{1},
\end{eqnarray}
on both equations in \eqref{dplusphi2}, together with
\eqref{dplusphi}, we get the spin three components $C^{3}_{\pm1}$ in terms of:
\[\left(\partial_{+}C^1_0, \ \partial_{-}C^1_0, \ \partial_{+} C^2_0, \ \partial_{-}C^2_0  \right),\] as shown in \eqref{C31}.

Now we  show how this process of reduction works at any spin level $s$. First we remind some useful properties of the lonestar product. Let us start by the generic product \[V^{s_1}_{m_1}\star V^{s_2}_{m_2}, \] that will reduce to a combination of the form:
\begin{equation}
V^{s_1+s_2-1}_{m_1+m_2}+\ldots+V^{s_1+s_2-1-j}_{m_1+m_2}+\ldots+V^{|m_1+m_2|+1}_{m_1+m_2},\label{prod}
\end{equation}
where we are not paying attention to the specific coefficients, which will be used in due time. 
The index $j$ goes from $0$ to $s_1+s_2-2-|m_1+m_2|$.
From (\ref{prod}) it follows that  the products:
$V^{s_1}_{m_1}\star a$ and $ a\star V^{s_1}_{m_1}$, with
$a=V^{2}_1-V^{2}_{-1}$, will contain combinations of the form:
\begin{equation}
V^{s_1+1}_{m_1+1}+V^{s_1+1}_{m_1-1}+\ldots,
\end{equation}
where the $\dots$ stand for lower total spin $s$ contributions. For our purposes only the highest total spin generators are relevant.

Furthermore,  for any chain of $2s-1$ generators with even spin $2s$  and even projections, $\sum\limits^{s-1}_{m=-s+1} V^{2 s}_{2m}+\ldots$,
further left or right multiplication by $a$ will change it into a chain of $2s$ generators
$\sum\limits^{s-1}_{m=-s} V^{2 s+1}_{2 m+1}+\ldots $
at the next spin level $2s+1$. As a consequence, arbitrary powers of
$a$ look like:
\begin{eqnarray}\label{a2s}
a^{2s}=\sum^s_{m=-s} V^{2 s+1}_{2 m}+\ldots \, \text{ and} \, \ a^{2s+1}=\sum^s_{m=-s-1} V^{2 s+2}_{2 m+1}+\ldots.
\end{eqnarray} 

From (\ref{C10}) and \eqref{masterfbtz}, it follows that
each $\partial_\pm$ derivative acting on $C^{1}_{0}$ is equivalent
to a left or right multiplication by $- a_{\pm\rho}$ inside the
trace. In particular,  taking $2s$ of these derivatives on $C^1_0$ is
equivalent  to take $2s$ powers of $\pm a_{\pm\rho}$ inside the
trace. 

The number of different derivatives of order $2s$
denoted by:  $\partial^{2s}_{\pm}$  is  $2s+1$. This number 
coincides precisely with  the number of  components with total spin=$2s+1$ in the
first power of \eqref{a2s}. So  one can use the $2s+1$ relations:
\begin{equation}\label{eqpmBTZ}
\partial^{2s}_{\pm} C^1_0=(a_{\pm \rho}^{2s}C)\big|_{V^1_0},
\end{equation}
to solve for  $2s+1$ components of
$C$:
\begin{equation}\label{eq:latticeIeven}
[C^{2s+1}_{2m}] \text{ with } m=-s, \ldots, s,
\end{equation}
in terms of components with lower total spin and their $\pm$ derivatives.  

One can always solve equations (\ref{eqpmBTZ})  in terms of (\ref{eq:latticeIeven}) because the set of symmetrised  powers   of
$a_{\pm\rho}^{2s}$ (more precisely,  their components with the highest total spin) will generate a basis for the $2s+1$ dimensional space generated by:
\[[V^{2s+1}_{2m}] \text{ with } m=-s-1, \ldots, s.\]
In order to prove this statement, we take the large $\rho$
limit. In this limit a given symmetric product $a^{2s}_\pm$ with
$2m_+$ plus signs and $2m_-=2\left(s-m_+\right)$ minus signs reduces
to a single basis element $V^{2s}_{2\left( m_+-m_-\right)}$. So, the set of all possible symmetric products
$a^{2s}_\pm$ span an $2s+1$-dimensional vector space. Consequently  the system of equations \eqref{eqpmBTZ} is  non-degenerate. 
 
Similarly, increasing the spin by one,  one can solve the $2s+2$ relations:
\begin{equation}
\partial^{2s+1}_{\pm}C^1_0=-(a_{\pm \rho}^{2s+1}C)\big|_{V^1_0},
\end{equation}
for the  $2s+2$ components 
\begin{equation}\label{eq:latticeIodd}
[C^{2s+2}_{2m+1}] \text{ with }m=-s-1,\ldots, ~s, 
\end{equation}
in terms of lower spin components and their $\pm$ derivatives.

Summarizing,  what we  have done is to  use the identities:
\begin{equation}\label{bijectionBTZ}
\partial_+ =- a_{\rho}\star_L, ~~~~~ \partial_- = - a_{-\rho}\star_R,
 \end{equation}
with left $\star_L$ and right $\star_R$ multiplication inside any trace. Notice that in Fourier space $(-i \partial_t ,-i \partial_\phi)=(w, k)$ the master field $(\ref{masterfbtz})$ is an eigenstate of the operators on the right hand side of (\ref{bijectionBTZ}). This will turn out to be a crucial observation, and it will be useful for later purposes,  but for now we just use (\ref{bijectionBTZ}) to solve for every component of $C^s_{m_s}$ with $(s,m_s)$ being points in a ``semi-lattice" with origin  $(1,0)$ and generated by positive integral combinations of basis vectors $(2,1)$ and $(2,-1)$. From now on we will refer to this particular ``semi-lattice" as $I$ and to the corresponding set of components of the master field $C$ in it as $C^{I}$.

 
 In exactly the same manner one can show how the set of powers 
 \begin{equation}\label{powers}a^{s_+}_{\rho}V^{2}_0 a^{s_-}_{-\rho},\end{equation}
 with $s=s_++s_-+1$ spans the complementary ``semi-lattice" of spin $s+1$ and
 projection $m_s=-s+1,-s+3,\ldots, s-3,s-1$ generators. Namely the ``semi-lattice" with origin at $(2,0)$ and positive integral combinations of  $(2,1)$ and  $(2,-1)$.
 We refer to it as $II$,  and the corresponding components of the master field $C$,
   $C^{II}$. More in detail, this means that we can solve the $s$ relations:
 \begin{equation}\label{latticeII}
 \partial_{+}^{s_+}\partial_{-}^{s_-} C^2_0=(-1)^{s_+ +s_-  }\left(a^{s_-}_{-\rho}V^{2}_0a^{s_+}_{\rho} C\right) \big|_{V^1_0},
 \end{equation}
for the set of components in $C^{II}$ with highest spin= $s+1$ and projections $m_s=-s+1,-s+3,\ldots, s-3,s-1$.

 \begin{itemize}
 \item In conclusion, equations \eqref{eqpmBTZ}-\eqref{eq:latticeIodd} and \eqref{latticeII} allow to solve for every components of $C^{I}$ and $C^{II}$ in terms of $C^1_0$ and $C^2_0$ and their derivatives along $\pm$ directions.\\
\end{itemize}

Finally,  the $V^{1}_0$-$d\rho$ component of \eqref{EqFluc} gives $C^2_0\sim \partial_\rho C^1_0$ and the  $V^{2}_0$-$d\rho$ component of \eqref{EqFluc} will determine the differential equation $D_2 C^1_0=0$ with 
\begin{equation}
D_2=\Box -\left(\lambda^2-1\right),\label{BTZLap}
\end{equation}
being the Klein Gordon operator in the BTZ$_{M=-1}$ background,  for a scalar field with mass squared $\lambda^2-1$. 

Now we go back to our case $\nu_0=\frac{1}{2}$ $\nu_{i>0}=0$. Here the $t$ component of
\eqref{EqFluc} is the same as for the BTZ$_{M=-1}$ case and so we use it as before
\begin{equation}
\partial_t C^{s-1}_{m_s+1}=C^{s}_{m_s}+C^{s}_{m_s+2}+\ldots,
\end{equation}
to solve for the highest spin,  with the lowest spin projection components
$(s,m_s)$. 
The dots  refer to  components with lower total  spin and we have omitted precise factors. That is, we solve
for all components in $C^{I}$ and $C^{II}$ in
terms of the line of highest weight and its contiguous
next-to-highest weight components, namely:
\begin{equation}\label{seed2}
C^{s+1}_{s} \text{ and } C^{s+2}_{s} \text{ with } s=0,\ldots,\infty.
\end{equation}
Next,  $\partial_\phi \sim a^{1+\tilde{s}_{Max}}+\text{ lower powers}$, and therefore from (\ref{a2s}) one can prove that the use of the $d\phi$ component of the equations \eqref{EqFluc} reduces 
the set of independent elements in \eqref{seed2} to:
\begin{equation}\label{seed3}
C^{s+1}_{s} \text{ and } C^{s+2}_{s} \text{ with }  0\leq s\leq s_{max},
\end{equation}
with $s_{max}+1$ being at  most $\tilde{s}_{max}+1$, the maximum value of the power in the polynomials $(P(a),\bar{P}(a))$, that determines the $\phi$ component of the connections ($A_\phi,\bar{A}_\phi$).
Notice that for some configurations in (\ref{backgrounds}) there are degeneracies and the number of independent components decreases in those cases. In fact $s_{max}$ determines the degree of the differential equation for $C^{1}_0$ (or equivalently
the number of $\rho$-components  one has to use to close
the system) to be given by $2\left(s_{max}+1\right)$, after the $\rho$
components of the equations of motion are imposed. \\ 
 
 \subsection {Solving the matter equations of motion}\label{sec:ChiDef1}
 

 In this subsection we  show how to proceed  for the simplest cases, and later on we prove in general that the equations of motion for scalars in (\ref{backgrounds}), can be expressed  in terms of simpler building blocks. 
Let us start by explicitly exhibiting  the solutions for matter fluctuations in the case of the backgrounds with $\mu_3\neq0$. Firstly, we determine the differential equation for $C^1_0$ by using the procedure outlined in the last paragraph of the previous section.
In this case $s_{max}=1$ and we get a differential equation for $ C^{1}_0$ with degree $2(s_{max}+1)=4$ in $\rho$. It is convenient to Fourier transform from $(\phi,t)$ to  $(k,\omega)$ for the  fileds
$C^s_m$ :
 \begin{equation}\label{eq:four}
 C^s_m[\rho,t,\phi]=e^{i \omega t } e^{i k \phi }C^s_m[\rho].
 \end{equation} 
The final form of the equation for $C^1_0$ is given in (\ref{D4}), here we will be somewhat schematic. After the change of coordinates $\rho=\tanh^{-1}{\left(\sqrt{z}\right)}$\footnote{Notice that this implies that $z$ lies in the positive real axis. The coordinate $z$ used in this section, should not be confused with the dynamical critical exponent $z$ introduced before, below equation (\ref{eq:lifini}). Our apologies for the confusion that this abuse of notation could create.} and the following redefinition of the dependent variable $C[z]=z^{\frac{-i \omega}{2}}(1-z)^{\frac{1-\lambda} {2}}G[z]$ one gets a new form for the original differential equation:
\begin{equation}
D_4 G[z]=0.
\end{equation}
The differential operator $D_4$, whose precise form is given in (\ref{D4}), has three regular singularities at  0,1 and $\infty$ with the following  $4\times 3=12$ characteristic exponents:
\begin{center}
\begin{tabular}{ccc}
$\alpha^{I}_0=(0,i \omega)$ &  $\alpha^{I}_1=(0,\lambda)$ & $\alpha_\infty=(\delta^+_-,\delta^{+}_+)$ \\
$\alpha^{II}_0=(1, 1+i \omega)$ & $\alpha^{II}_1=(1,1+\lambda)$ & $\widetilde{\alpha}_\infty=(\delta^{-}_-,\delta^{-}_+)$,
\end{tabular}
\end{center}
where:
\begin{eqnarray}
\delta^{+}_{+}&=\frac{1-\lambda}{2}+\delta^{+}_0(\mu_3) ,\ \ \delta^{+}_{-}=\frac{1-2 i \omega-\lambda}{2}-\delta^{+}_0(\mu_3), \nonumber   \\
\delta^{-}_{+}&=\frac{1-\lambda}{2}+\delta^{-}_0(\mu_3), \ \ \delta^{-}_{-}=\frac{1-2 i \omega-\lambda}{2}-\delta^{-}_0(\mu_3),
\end{eqnarray}
and:
\begin{eqnarray}\label{delta0}
\delta^{\pm}_0(\mu_3)&=\frac{-3 \pm \sqrt{9-36 i \mu_3(\omega+k)+12\mu_3^2(\lambda^2-1)}}{12 \mu_3}.
\end{eqnarray}
Notice that $\delta^{+}_0$ is regular in the limit of vanishing $\mu_3$ whereas $\delta^{-}_0$ is not.

For a Fuchsian differential equation of order $n$ with $m$ regular singular points
the sum of characteristic exponents is always
$(m-2)\times\frac{n(n-1)}{2}$ \citep{Ilyashenko+Yakovenko}. It is easy
to check that in our case $n=4$, $m=3$ the sum of characteristic
exponents is indeed $6$. An interesting case is when $n=2$ and $m=3$
in that case one has $m \times n=6$ characteristic exponents whose sum
equals $1$. Conversely, it is a theorem that any set of $6$ numbers
adding up to $1$ defines a unique Fuchsian operator of order $n=2$
with $m=3$ regular singular points. It is also a theorem that such a sextuple of roots defines a subspace of solutions that carry an irreducible representation of the monodromy group of $D_n$ and hence a factor $D_2$ \citep{Ilyashenko+Yakovenko}. Namely:
\begin{equation}
D_n=D^L_{n-2}D^R_{2}, \label{factor}
\end{equation}
and $D^L_{n-2}$ is also Fuchsian and the $L$ and $R$ denote the left
and right  operator, respectively, in the factorisation.

Before proceeding, let us review some facts that will be
used in the following \citep{Ilyashenko+Yakovenko,abramowitz+stegun}. The most
general form of a Fuchsian  differential operator $D_2$ once the position of the regular singular points are fixed  at $0,1,\infty$ and a pair of characteristic exponents is fixed to zero, is: 
\begin{equation}\label{D2}
D_2\equiv y(y-1)\frac{d^{2}}{dy^2}+\left((a+b+1)y-c\right)\frac{d}{dy}+ab.
\end{equation}
The characteristic exponents are:
\begin{eqnarray}\label{cexponents}
\alpha_{0}=(0,1-c), \ \alpha_1=(0,c-a-b), \ \alpha_\infty=(a,b).
\end{eqnarray}
The kernel of $D_2$ is generated by the linearly independent functions:
\begin{eqnarray}
u_1(a,b,c | z)&\equiv & \, _2F_1(a,b,c | z),\nonumber\\
z^{1-c}u_2(a,b,c | z)&\equiv  & z^{1-c} \, _2F_1(a+1-c,b+1-c,2-c | z),\label{flucBTZ}
\end{eqnarray}
which are eigenstates of the monodromy action at $z=0$. The second solution is independent only when $c$ is not in $\mathbb{Z}$.  The monodromy eigenstates at $z=1$ are:
\begin{eqnarray}
\tilde{u}_1(a,b,c | z)&\equiv & \, _2F_1(a,b,1+a+b-c |1- z),\nonumber\\
(1-z)^{c-a-b}\tilde{u}_2(a,b,c | z)&\equiv  & (1-z)^{c-a-b} \, _2F_1(c-a,c-b,1+c-a-b |1- z),\nonumber\\\label{flucBTZ2}.
\end{eqnarray}  
when $c-a-b$ is not in $\mathbb{Z}$. In a while we will see that $c-a-b=\lambda$.

 Our operator $D_4$ does have the properties mentioned in the paragraph
 before \eqref{factor}. In fact each one of the set of characteristic exponents:
\begin{eqnarray}
& \left(\alpha^I_0, \, \alpha^I_1, \, \alpha_{\infty}\right),\nonumber \\
& \left(\alpha^I_0, \, \alpha^I_1, \, \widetilde{\alpha}_{\infty} \right),
\end{eqnarray}
adds up to 1, and hence defines the second order  Fuchsian operators:
\begin{eqnarray}\label{defHyper1}
D^R_2 &:& a= \delta^{+}_+(\mu_3),~ \, b=\delta^{+}_- (\mu_3), ~ \, c=1-i \omega, \nonumber    \\  \label{defHyper2}
\widetilde D^R_2 &:& a= \delta^{-}_+ (\mu_3),~ \, b=\delta^{-}_- (\mu_3),~ \, c=1-i \omega.
\end{eqnarray}
As a result $D_4$ has two independent factorizations:
\begin{eqnarray}
D_4=D^L_2 D^R_2 \text{ and } D_4=\widetilde{D}^L_2 \widetilde{D}^R_2,
\end{eqnarray}
as one can check explicitly. Consequently we have:
\begin{equation}
ker D_4=ker D^{R}_2 \bigoplus ker \widetilde{D}^{R}_2,
\end{equation}
where  $ker D^{R}_2$ is given by  the   hypergeometric functions $u_1$ and $u_2$ given in \eqref{flucBTZ}, with the parameters $a$, $b$ and $c$ defined in \eqref{defHyper1}. 
This proves that the fluctuation equation in the background $\mu_3\neq0$  is solved in terms of  four linearly independent hypergeometric functions, which, from now on we refer to as  ``building blocks".  

One can explicitly verify this factorization pattern  for the next background, with $\mu_3, \mu_5\neq 0$. In this case $s_{Max}=3$ and the corresponding differential operator $D_8$, has order $8$, and is again Fuchsian with $3$ regular singularities in the $z$ coordinate system previously defined (we always place them at 0, 1 and $\infty$). The characteristic exponents are:

\begin{center}
\begin{tabular}{ccc}
$\alpha^{I}_0=(0,i \omega)$ &  $\alpha^{I}_1=(0,\lambda)$ & $\alpha^{I}_\infty=(\delta^{++}_-,\delta^{++}_+)$\\
$\alpha^{II}_0=(1, 1+i \omega)$ & $\alpha^{II}_1=(1,1+\lambda)$ & $\alpha^{II}_\infty=(\delta^{+-}_-,\delta^{+-}_+)$ \\
$\alpha^{III}_0=(2,2+i \omega)$ &  $\alpha^{III}_1=(2,2+\lambda)$ & $\alpha^{III}_\infty=(\delta^{-+}_-,\delta^{-+}_+)$\\
$\alpha^{IV}_0=(3,3+i \omega)$ & $\alpha^{IV}_1=(3,3+\lambda)$ & $\alpha^{IV}_\infty=(\delta^{--}_-,\delta^{--}_+)$,
\end{tabular}
\end{center}
where for each of the couples of exponents $\alpha_\infty $ the
following property holds: 
$\delta^{\pm\pm}_{+}(\mu_3,\mu_5)+\delta^{\pm\pm}_{-}(\mu_3,\mu_5)=1-i \omega-\lambda $. As a  consequence there are four triads of characteristic exponents whose sums equal 1 :
\begin{eqnarray}\nonumber
& \left(\alpha^I_0, \, \alpha^I_1, \, \alpha^{I}_{\infty}\right),  & \left(\alpha^I_0, \, \alpha^I_1, \, \alpha^{II}_{\infty} \right),\\
& \left(\alpha^I_0, \, \alpha^I_1, \, \alpha^{III}_{\infty}\right), & \left(\alpha^I_0, \, \alpha^I_1, \, \alpha^{IV}_{\infty} \right).
\end{eqnarray}
Each  of them defines a second order  ``Hypergeometric operator'' as in
\eqref{defHyper1}:
\begin{eqnarray}\nonumber
D^{I\, R}_2, D^{II\, R}_2, \, D^{III\, R}_2 \text{ and } D^{IV\, R}_2
\end{eqnarray}
such that
\begin{equation}\nonumber
ker D_8=ker D^{I \, R}_2 \bigoplus ker D^{II \, R}_2 \bigoplus ker D^{III \, R}_2\bigoplus ker D^{IV \, R}_2.
\end{equation}

In fact there is a simple way to prove that  the above pattern generalises, showing that 
the solutions of our higher order differential equations can be expressed in terms of ordinary hypergeometric functions,  
for all of the representatives in (\ref{backgrounds}). The
point is to use the fact that the Fourier components $C(\omega,k)$ of the full
master field $C(t,x)$ defined by the arbitrary polynomial $P_\phi$ and
$\bar{P}_\phi$, are eigenstates of the operators in the right hand
side of: \\
\begin{eqnarray}
\partial_t&=&\frac{-a_\rho\star_L + a_{-\rho}\star_R}{2},\nonumber\\
\partial_\phi&=&-\left(\frac{a_\rho}{2}+P_{\phi}(a_\rho)\right)\star_L-\left(\frac{a_{-\rho}}{2}-\bar{P}_\phi(a_{-\rho})\right)\star_R ,  \label{Opld}
\end{eqnarray}
\\
with eigenvalues $(i \omega,i k)$ respectively. The same can be said of the trace component $C^1_0(\omega,k)$ but in this case, the left and right multiplication are equivalent by cyclic property of the trace. As the operators on the
 right hand side of  (\ref{Opld}) are polynomials in  $a_{\pm \rho}$, they share eigenvectors with the latter.  But as we pointed out around (\ref{bijectionBTZ}):
\begin{eqnarray}
i(\omega^\prime+k^\prime) C_{BTZ}(\omega^\prime,k^\prime)&=&-a_{\rho}\star_L C_{BTZ}(\omega^\prime,k^\prime),\nonumber\\
i(k^\prime-\omega^\prime)C_{BTZ}(\omega^\prime,k^\prime)&=&-a_{-\rho}\star_R C_{BTZ}(\omega^\prime,k^\prime),\label{OpldBTZ}
\end{eqnarray}
where $C_{BTZ}$ is the master field for the BTZ$_{M=-1}$ connection. So from (\ref{Opld}) and (\ref{OpldBTZ}) it follows that:
\begin{eqnarray}
C^1_0(\omega,k)=C^1_{0 BTZ}(\omega^\prime,k^\prime), \label{Integrability}
\end{eqnarray}
where $(\omega^\prime,k^\prime)$ are any of the roots of the algebraic equations:
\begin{eqnarray}
i \omega&=& i\omega^\prime,\nonumber \\
i k&=& ik^\prime-\left(P_\phi(-i(\omega^\prime+k^\prime))-\bar{P}_\phi(-i(k^\prime-\omega^\prime))\right). \label{secular}
\end{eqnarray}
Relations \eqref{Integrability} imply that the differential equation for $C^{1}_0$ in the class of ans\"atze \eqref{backgrounds} is always integrable in terms of hypergeometric functions $_2F_1$. The number of linearly independent modes being given by twice the order of the algebraic equations \eqref{secular}, which can be checked to be, $2(s_{Max}+1)$. Here $s_{Max}+1$ coincides with the order of the polynomial equation (\ref{secular}) for $k^\prime$ in terms of $(\omega,k)$.

Summarizing, the most general solution for fluctuations in (\ref{backgrounds}) is:
\begin{align}
 \quad \quad \quad \quad \quad C^1_0(\omega,k)= \sum_r e^{i \left(\omega t + k \phi \right)}(1-z)^{\frac{1-\lambda}{2}}&\left(c^{in}_r z^{ - \frac{i \omega}{2}} u_1(a_r,b_r,1-i \omega, z) \right. \nonumber\\&+\left. c^{out}_r z^{\frac{ i \omega}{2}} u_2(a_r,b_r,1-i \omega, z)\right),\nonumber\\ 
  a_r \equiv \frac{i(k^\prime_r-\omega)+1-\lambda}{2}&, ~ b_r\equiv\frac{-i(k^\prime_r+\omega)+1-\lambda}{2}, ~~~~~~~~~~~~~~~~~~~~~~~~~~ \label{generalsol}
\end{align} 
where $k^\prime_r$ are the roots of \eqref{secular} and $r=1,\ldots,2(s_{Max}+1)$. 

For later reference we write down \eqref{generalsol} in terms of monodromy eigenstates at the boundary $z=1$:
\begin{eqnarray}
 C^1_0(\omega,k)&=& \sum_r e^{i \left(\omega t + k \phi \right)}z^{\frac{-i \omega}{2}}(1-z)^{\frac{1-\lambda}{2}}\left(\tilde{c}^1_r \tilde{u}_1(a_r,b_r,1-i \omega; z)\right. \nonumber\\&&~~~~~~~~~~~~~~~~~~~~~~~~+\left. \tilde{c}^{2}_r(1- z)^{\lambda} \tilde{u}_2(a_r,b_r,1-i \omega; z)\right). \label{generalsol2}
\end{eqnarray} 
As a  check, let us reproduce the first result of this section by using this method. For the case $\mu_3\neq0$ the equation for $k^\prime_r$ are:
\begin{equation}
i k=i k^\prime_r-\mu_3\left(-(\omega+k^\prime_r)^2+\frac{1-\lambda^2}{3}\right),
\end{equation}
whose solutions are :
\begin{equation}
i k^\prime_\pm=- i \omega  -\delta^{\pm}_0(\mu_3).
\end{equation}
This coincides with the solution one obtains from \eqref{defHyper1}, as can be seen using the definitions in the second line of \eqref{generalsol}. We note that only $k^\prime_+$ is smooth in the BTZ limit $\mu_3$ to zero.

As an interesting observation, we would like to draw the attention of the reader to the fact that the boundary conditions
for the most general fluctuation (\ref{generalsol}) at the horizon and
boundary, $z=0$ and $z=1$, respectively, are not affected by the fact that connections \eqref{backgrounds} and  the corresponding background tensor
fields $g^{(n)}$, defined as (\ref{tensor}), do break the original $BTZ_{M=-1}$ boundary conditions. \\

\section{QNM and bulk to boundary 2-point functions} \label{sec:QNM}
As anticipated, in this subsection we will further argue that  the connections (\ref{backgrounds}) describe a class of black hole configurations. 
We will do so by showing the presence of quasi normal modes (QNM). We will compute their spectrum for any representative in 
(\ref{backgrounds}) and, in particular, more  explicitly for the simplest cases discussed in the previous section. 

We start by recalling the conditions for quasi normal modes in AdS  black
holes \citep{Horowitz:1999jd}: they behave like ingoing waves at the horizon, $z=0$  and as subleading
modes at the boundary $z=1$. In the language employed before, the relevant conditions
reduce to ask for solutions with indicial roots $\alpha_0=0$ at the
horizon $z=0$, and $\alpha_1=\lambda$ at the boundary $z=1$.
In this section we are considering the region $0<\lambda<1$ so
that $(1-z)^{\frac{(1-\lambda)}{2}}$ is the leading behaviour near
the boundary. In terms of the most general solution
\eqref{generalsol}, the ingoing wave condition reads:
$c^{out}_r=0$. The subleading behaviour requirement implies the
quantisation conditions\footnote{ We have the identity $_2F_1[a,b,c,z]=\frac{\Gamma [c]\Gamma [a+b-c]}{\Gamma [c-b]\Gamma [c-a]} \, _2F_1[a,b,a+b-c+1,1-z]+(1-z)^{c-a-b}\frac{\Gamma [c]\Gamma [c-a-b]}{\Gamma [b]\Gamma [a]} \,_ 2F_1[c-a,c-b,c-a-b+1,1-z]$ \citep{abramowitz+stegun}. The quantisation condition \eqref{quantisationCond} is equivalent to $c-a=-n$ and $c-b=-n$ respectively. These choices guarantees 
that the first term on the rhs of the previous identity vanishes. Indeed, this is the term that carries the leading behaviour of the field at the boundary.}.
\begin{equation}
\omega \pm k^\prime_r+ i (1+2 n+\lambda)=0, ~~ r=1,\ldots2(s_{Max}+1),\label{quantisationCond}
\end{equation}
where $n$ is an arbitrary and positive integer.

We should elaborate about the smoothness of the quasi normal modes at the horizon. In the Eddington-Finkelstein coordinates $v=t-\frac{ T}{2} log(\rho)+\ldots$ and $\rho^*=\frac{4}{T^2} \rho^2+\ldots$, see \eqref{lineElement} the incoming waves, namely the $c^{in}_r$ modes , behave as plane waves $e^{I w v}$, at leading order in the near-horizon expansion. In contrast, the $c^{out}_r$ modes are not $C^\infty$ as they look like $e^{i \omega v}\left({\rho^*}^{i \omega}\right)$. In other words, the requirement of incoming waves at the horizon  amounts to have a  smooth solution at the horizon \citep{Horowitz:1999jd}. 

 In our example $ \mu_3\neq0$, $s_{Max}=1$, there are $2\times2$ branches in the quantisation conditions \eqref{quantisationCond}. The associated branches of quasi normal modes being:
 \begin{eqnarray}\small
 \omega^0_{n}&=&-k-i\left(1+2n+\lambda-\frac{2\mu_3}{3}\left(1+(1+2\lambda)(1+\lambda)-\lambda^2+6 n(1+\lambda) +6 n^2\right)\right),\nonumber \\\omega^{\pm}_{n}&=&-\frac{1}{2}i (1+2 n +\lambda)+ \delta^{\pm}(n,\mu_3),~~~~~~~~~~~~~~~~~~~~~~~~~~~~~~~~~~~~~~~~~~~~~~~~~~\label{QNM}
 \end{eqnarray}
 where:
 \begin{equation}\small
 \delta^\pm(n,\mu_3)=\frac{-i \pm\sqrt{-1+8 (1+2 i k+2 n+\lambda ) \mu _3-\frac{16 \left(\lambda ^2-1\right) \mu _3^2}{3}}}{8 \mu_3}.
 \end{equation}
Before going on,  let us briefly mention some relevant issues about the stability of the branches \eqref{QNM}.  It is not hard to see that for large enough values of $k\in \mathbb{R}$ at least one of the branches $\omega^\pm_n$ will exhibit a finite number of undamped modes, namely modes with positive imaginary parts. However for a fixed value of $k$ and $\mu_3$ the UV modes ($n\gg 1, k, \mu_3$) will go like $\omega^\pm_n \sim  -i~ n $ and hence will be stable. The branch $\omega^0_n$ is stable for $\mu_3<0$.
Finally notice also that $(\omega^0_n,\omega^+_n)$ become the left and right moving branches of the BTZ$_{M=-1}$ case, in the limit of vanishing $\mu_3$, whereas $\omega^-_n$ is not analytic in that limit. 
 
 We  have $2\times2(s_{Max}+1)$ independent  solutions $(c^{in},c^{out})_r$ in \eqref{generalsol}. Each block $r$ represents an independent degree of freedom and a general fluctuation in the background (\ref{backgrounds}) can be re-constructed as a  combination of them. So, for the moment we restrict our analysis to a given sector, let us say the block $r$.

In order  to define the bulk to boundary 2-point function we set  $\tilde{c}^2_r=0$ in \eqref{generalsol2}, corresponding  to  the solution with 
the leading behaviour $(1-z)^{\frac{1-\lambda}{2}}$ at the boundary. We will further  fix $\tilde{c}^1_r =1$, to guarantee independence  on $\omega$ and $k$ of the leading term
in the expansion of the solution near the boundary, in such a way that its Fourier transform becomes proportional to $\delta^{(2)}(t,\phi)$ at the boundary, 
which  is the usual UV boundary condition in coordinate space. As a result, in Fourier space, the bulk to boundary 2-point function of the block of solutions $r$ is given by:
 \begin{equation}
G^{(2)}_r(\omega,k,z)\equiv  \tilde{u}_1(a_r,b_r,1-i \omega;1- z). \label{2pointrFourier}
 \end{equation}
After  Fourier transforming back  in $(t,\phi)$ space and using the  $\rho$ coordinate one gets preliminary:
\begin{equation}
G^{(2)}_r(t,\phi,\rho)=J_r(-i \partial_t,-i \partial_\phi) \left(G^{(2)}_{BTZ}(t,\phi;\rho)+\delta G^{(2)}_r(t,\phi,\rho)\right). \label{2pointr}
\end{equation}
 We stress that (\ref{2pointr}) obeys the boundary condition:
\begin{equation}
G^{(2)}_r(t,\phi,\rho)\rightarrow \delta^{(2)}(t,\phi), \text{ when $\rho\rightarrow \infty$.} \label{bdrycond}
\end{equation}
The quantity:
\begin{equation}
J_r(\omega,k)\equiv \frac{1}{\frac{\partial k^{\prime}_r(\omega,k)}{\partial k} } e^{i\big(k-k^{\prime}_r(\omega,k)\big)\phi}, \nonumber
\end{equation}
is the product of the Jacobian from the change of variables from $k$ to $k^{\prime}_r$ times an exponential contribution. For our specific  case:
\begin{equation}
J_r(\omega,k)=\left(1+2i\mu_3 \delta^{\pm}_0(\omega,k)\right)  e^{i\big(k-k^{\prime}_r(\omega,k)\big)\phi}. \label{ex1}
\end{equation}
The quantity:
\begin{equation}
G^{(2)}_{BTZ}(t,\phi,\rho)=-\frac{\lambda}{\pi}\left(\frac{e^{-\rho}}{e^{-2\rho}\cosh{x_+}\cosh{x_-}+\sinh{x_+}\sinh{x_-}}\right)^{1-\lambda},  \label{2pointbtz}
\end{equation}
is the bulk to boundary 2-point function for BTZ$_{M=-1}$. Notice that \eqref{2pointbtz} is smooth in the near-horizon expansion as its leading contribution is independent of $t$. We note that the contributions coming from $G^{(2)}_{BTZ}$ to \eqref{2pointr} are also smooth at the horizon provided the Taylor expansion of $J_r(w,k)$ around $(\omega,k)=0$ starts with a constant or an integer power of $k$. This is always the case, as one can infer from \eqref{secular} that $J_r=1+O(\mu_3)$, as in the particular case (\ref{ex1}).

Finally $\delta G^{(2)}_r$ is a contribution that comes from the deformation of the countour of integration that follows from the change $k\rightarrow k^\prime_r$ . The change of variable from $k$ to $k^\prime_r(\omega,k)$ deforms the real line $\mathbb{R}$ to a contour $C_{r,\omega}\equiv k_r^\prime(\mathbb{R},\omega)$. Integration over the contours $k^\prime_r \in \mathbb{R}$ and $k^\prime_r \in C_{r,\omega}$ (followed by integration over $\omega \in \mathbb{R}$) of the integrand
\begin{equation}\nonumber
e^{i k^\prime_r \phi+ i \omega t}  \tilde{u}_1(a_r,b_r,1-i \omega;1- z),
\end{equation}
differ by the quantity $\delta G^{(2)}_r(t,\phi,z)$.  This quantity can be obtained imposing the condition (\ref{bdrycond}). In Fourier space $(\omega,k^\prime_r)$ It reads:
\begin{equation}
\delta G^{(2)}_r(\omega, k^\prime_r,z)=\left( \frac{\partial k^{\prime}_r}{\partial k}-1\right) \tilde{u}_1(a_r,b_r,1-i \omega;1- z).\footnote{Notice that the quantity $\delta G^{(2)}_r(\omega,k^\prime_r, z)$ ( as $G^{(2)}_{BTZ}(\omega, k^\prime_r,z)$) is in the kernel of the BTZ Klein-Gordon operator $D_2(\omega,k^\prime_r,z)$.}
\end{equation}
Finally, (\ref{2pointr}) takes the form:
\begin{equation}
G^{(2)}_r(t,\phi,\rho)= e^{{-\big(i k^{\prime}_r\left(-i \partial_t,-i \partial_\phi\right)- \partial_\phi\big)\phi}} G^{(2)}_{BTZ}(t,\phi,\rho).\footnote{We note that the $\phi$ in the exponential \eqref{G2final} is located to the right of the derivatives.}\label{G2final}
\end{equation}
 For the same reasons explained before (\ref{G2final}) is smooth at the horizon, namely its leading behaviour is independent on $t$.



Notice that periodicity under $t\rightarrow t+2\pi i$ is preserved by all building blocks \eqref{G2final}. The preservation of thermal periodicity comes after imposing the BTZ holonomy condition on \eqref{backgrounds}. It is a global statement in the sense that is determined by the exponentiation properties of the algebra. Namely the gauge group elements generating the family \eqref{backgrounds} with $\nu_0=\frac{1}{2}$, $\nu_{i>0}=0$:
\begin{eqnarray}
g&=&e^{-\rho V^2_0}e^{-\frac{a}{2} t-\left(\frac{a}{2}+P_\phi(a)\right)\phi}, \nonumber\\
\bar{g}&=&e^{\rho V^2_0}e^{-\frac{a}{2} t+\left(\frac{a}{2}-\bar{P}_\phi(a)\right)\phi}, \label{conical}
\end{eqnarray}
are thermal periodic due to the fact $i \pi a $ exponentiates to the center of the group whose Lie algebra is $hs(\lambda)$ \citep{Monnier:2014tfa}.


\subsection{Making contact with other relevant backgrounds}\label{sec:ChGau} 
 In this section we perform the gauge transformations \eqref{LKGP} and \eqref{LBT} taking our backgrounds  to the GK (BHPT2) ones.
 As already said, the backgrounds to be transformed have critical exponent $z<1$. Here we will focus in performing gauge transformations (\ref{LKGP}) and (\ref{LBT}) on the scalar fluctuations for $\bar{\mu}_3=-\mu_3\neq0$ and we will explicitly verify that they solve the equation of motion for matter  fluctuations in the GK  (BHPT2) backgrounds. The analysis will be done perturbatively, to first order in  a $\mu_3$ expansion.
 
  To this purpose we introduce the series expansion:
  \begin{equation}
 C=\sum_{i=0}^\infty \mu_3^i ~\overset {(i)}{C},
 \end{equation}
 for the master field in equations \eqref{EqFluc} with the connections $(A,\bar{A})$ given by \eqref{KGPBT}, \eqref{KGP} and \eqref{BT}. Taking the $\mu_3^i$ component of \eqref{EqFluc}:
 \begin{eqnarray}
 (d +\overset{(0)}{\mathcal{A}}\star_{L}-\overset{(0)}{\bar{\mathcal{A}}}\star_R)\overset{(i)}{C}=-\sum^{i}_{j=1}(\overset{(j)}{\mathcal{A}}\star_{L}-\overset{(j)}{\bar{\mathcal{A}}}\star_R)\overset{(i-j)}{C}, ~i=0,\ldots,~\infty, \label{eqG}
 \end{eqnarray}
where $\overset{(j)}{\mathcal{A}}$ is the coefficient of $\mu^j_3$ in the Taylor expansion of $\mathcal{A}$ about $\mu_3=0$. Notice that if $\overset{(i)}{C}$ is a particular solution of \eqref{eqG}, then $\overset{(i)}{C}+\text{constant}~ \overset{(0)}{C}$ is also a solution. This is in fact the maximal freedom in defining $\overset{(i)}{C}$ and it constraints the form of the ``folded"  version of (\ref{eqG}) to be of the form:
 \begin{eqnarray}
 D_2\overset{(0)}{C^1_0}&=&0,~ i=0,
 \nonumber\\
 D_2 \overset{(i)}{C^1_0}&=&\overset{(i)}{D}\left( \overset{\scriptsize(0)}{C^1_0},\ldots,\overset{\scriptsize(i-1)}{C^{1}_0} \right), ~i=1,\ldots\infty, \label{eqGUnfolded}
 \end{eqnarray}
where the differential operator $D_2$ is
the BTZ Klein-Gordon operator  \eqref{BTZLap} and $\overset{(i)}{D}$ is a linear differential operator in $\rho$ that we shall find out explicitly when analysing up to first order in $\mu_3$. 

Let us write down the connections  \eqref{backgrounds} with $\mu_3=-\bar{\mu}_3\neq0$ as:
\begin{equation}
\mathcal{A}_{ours}=\overset{(0)}{\mathcal{A}}+\mu_3 \overset{(1)}{\mathcal{A}}_{ours}, ~ \mathcal{A}_{ours}=\overset{(0)}{\mathcal{A}}+\mu_3 \overset{(1)}{\mathcal{A}}_{ours}.\label{c1}
\end{equation}
The full answer ${C^1_0}_{ours}$ is defined as the building block $r$ in \eqref{generalsol} with $k^\prime_r$, given by the root \eqref{kz<1} of equation \eqref{eqkz<1} which is the analytic solution in the limit $\mu_3$ to zero. By using the folding method one can check until arbitrary order in $i$ that \eqref{eqGUnfolded} works for the expansion coefficients $\overset{(i)}{C}_{ours}$. Here we restrict to  the $i=1$:
\begin{equation}
D_2\overset{(1)}{{C}^1_0}_{ours}=\overset{(1)}{D}_{ours}\overset{(0)}{C^1_0},\label{OursDiffEq}
\end{equation}
where:
\begin{equation}
\overset{(1)}{D}_{ours}=\frac{16 i k e^{2 \rho } \left(\frac{1}{3}(\lambda^2-1)+ k^2+ w^2\right)}{ \left(e^{2 \rho }+1\right)^2}.
\end{equation}
Let us  solve \eqref{OursDiffEq}. We can  expand in series the  solution for ${C^1_0}_{ours}$ (\ref{generalsol}), but we will use gauge covariance instead. From the use of the transformation laws:
\begin{eqnarray}
\mathcal{A}_{ours}=e^{\Lambda_{ours}}\mathcal{A}e^{-\Lambda_{ours}}+e^{\Lambda_{ours}}d~ e^{-\Lambda_{ours}}, \nonumber\\ \bar{\mathcal{A}}_{ours}=e^{\bar{\Lambda}_{ours}}\bar{\mathcal{A}}e^{-\bar{\Lambda}_{ours}}+e^{\bar{\Lambda}_{ours}}d~ e^{-\bar{\Lambda}_{ours}},
\end{eqnarray}
at linear order, with:
\begin{eqnarray}
\Lambda_{ours}=- \phi P_\phi(a_\rho), ~ \bar{\Lambda}_{ours}=- \phi \bar{P}_\phi(a_{-\rho}),
\end{eqnarray}
and ${C^1_0}_{ours}=\left((e^{\Lambda_{ours}- \bar{\Lambda}_{ours}})\overset{(0)}{C^1_0}\right)\bigg|_{V^1_0}$,  for the case $\mu_3=-\mu_3\neq0$ in Fourier space, it follows that:
\begin{eqnarray}
\overset{(1)}{C^{1}_0}_{ours}&=&-i \partial_k \left((a_\rho^2+a_{-\rho}^2-\text{ trace})\overset{(0)}{C} \right)\bigg|_{V^1_{0}}\nonumber\\&=&-i \left(\frac{2}{3}(1-\lambda^2)-2(k^2+w^2)\right)\partial_k \overset{(0)}{C^1_0} +\ldots, \label{C110ours}
\end{eqnarray}
where the $\ldots$  in (\ref{C110ours}) stand for terms that are proportional to $\overset{(0)}{C^1_0} $ and hence are in the kernel of $D_2$.
 
 To check that \eqref{C110ours} is solution of \eqref{OursDiffEq} it is enough to check that:
\begin{equation}
\bigg[i \left(\frac{2}{3}(1-\lambda^2)-2(k^2+w^2)\right)\partial_k,~D_2\bigg]=\overset{(1)}{D}_{ours},
\end{equation}
by using \eqref{D2rho} or to notice that (\ref{C110ours}) coincides with the first order coefficient in the Taylor expansion around $\mu_3=0$ of the corresponding  solution ${C^1_0}_{ours}$ which is given by $(\frac{\partial k^\prime}{\partial \mu_3}\partial_{k^\prime}{C^1_0}_{ours})|_{\mu_3=0} =\frac{\partial k^\prime}{\partial \mu_3}|_{\mu_3=0}\partial_{k}\overset{(0)}{C^1_0}$.

Next, we truncate the GK background at first order in $\mu_3$ and after following the procedure we can explicitly show again that the form \eqref{eqGUnfolded} holds until $i=1$ \footnote{We checked it up to $i=2$, when the GK background is truncated at second order in $\mu_3$.}. Here we just present the $i=1$ equation:
\begin{equation}
D_2 \overset{(1)}{C^1_0}_{GK}=\overset{(1)}{D}_{GK}\overset{(0)}{C^1_0}.\label{KGPDiffEq}
\end{equation} 
 The expression for $\overset{(1)}{D}_{GK}$ is given in \eqref{eq:D1KGP}. We should stress again that \eqref{KGPDiffEq} refers only to fluctuations over the GK ansatz that are analytic when $\mu_3$ goes to zero.  
Finally we check explicitly that the transformed fluctuation:
 \begin{eqnarray}
 \overset{(1)}{{C^{1}_0}}_{GK}&=&\overset{(1)}{C^1_0}_{ours}+\left((\overset{(1)}{\Lambda}_{GK}-\overset{(1)}{\bar{\Lambda}}_{GK})\overset{(0)}C\right)\bigg|_{V^1_0}\nonumber \\
 &=&\overset{(1)}{C^1_0}_{ours}-\frac{i k \left(3 e^{2 \rho }+5\right) }{3 \left(e^{2 \rho }+1\right)^2}\left(\left(e^{2 \rho }-1\right) \overset{(0)}{C^1_0}-\left(e^{2 \rho
   }+1\right)\partial_\rho\overset{(0)}{C^1_0}\right),
 \end{eqnarray}
 solves \eqref{KGPDiffEq}, after using \eqref{OursDiffEq} and the $i=0$ equation in (\ref{eqGUnfolded}). We have then reproduced the result of  \citep{Kraus:2012uf,Gaberdiel:2013jca},
 by starting  from our ansatz.\\

\section{Final remarks}

We have presented a family of connections constructed out of arbitrary polynomial combinations of the BTZ$_{M=-1}$ connection in $hs(\lambda)\times hs(\lambda)$ 3D CS theory. Their space time tensor fields present smooth horizons. The system of higher order differential equations of motion for matter fluctuations  can be solved in terms of hypergeometric functions related to the solutions in the  BTZ background. This allows to solve explicitly for Quasi Normal Modes and 2-point functions. As a check, we have made contact with other backgrounds  studied  in the literature. 
Among the open problems that this work leaves unanswered, we mention the following ones.  The first  regards the understanding of which (higher spin ?) charges are carried by these backgrounds, or, more generally
what is the asymptotic symmetry algebra associated to them. Recent progress on this problem for black holes in the $Sl(3)$ CS theory, may allow to get an answer for the cases presented here.
Secondly, one would like to use the results found here for the matter fluctuations, to solve for more general backgrounds by using the non residual gauge transformations that
carry our backgrounds to these.
Unfortunately, a perturbative analysis along the lines discussed in this paper seems to be unavoidably  beset by singularities at the horizon $\rho=0$. 
It would be interesting  to know whether this is an artifact of the perturbative expansion and if a full non perturbative analysis would be free of such singularities. This would allow to study quasi normal modes 
virtually for any black hole.

We owe a more detailed study of the properties of the differential operators governing the propagation of matter in the backgrounds here presented. Perhaps this study could shed some light on the specific geometrical properties that drive matter propagation in generic backgrounds with higher spins  \citep{Monnier:2014tfa}. Finally, we stress that the same approach we followed to show the factorisation property, can be implemented for a family of backgrounds constructed out of polynomials in more general highest weight connections. We hope to come back to some of these issues in the near future.  \\ \\

\paragraph{Acknowledgments} 
KSN acknowledges partial support from the European Commission  under contract  PITN-GA-2009-237920. ACB acknowledges hospitality and feedback received from the I.E. Tamm Department of Theoretical Physics of Lebedev Physical Institute, during the presentation of this work. We thank G. Mandal, R. Sinha and N.Sorokhaibam for private communication in relation to a typo in the previous version of the first line of equation (\ref{QNM}).
\appendix
\section{Conventions} \label{app:Conv}
The construction of the $hs(\lambda)$ algebra can be seen for example in \citep{Gaberdiel:2012uj}. The algebra is spanned by the set of generators $V^{s}_{t}$ with $s=2,\ldots, \infty$ and $1-s \leq t\leq s-1$. The element $V^1_0$ denotes the identity operator. To define the algebra we use the $\star$-product representation constructed in \citep{Pope:1990kc}:
\begin{equation}
\label{eq:lonestar}
V^{s}_{m}\star V^{t}_{n}=\frac{1}{2}\mathlarger{\mathlarger{\mathlarger{\mathlarger{\sum}}}}^{s+t-Max[\left|m+n\right|,\left|s-t\right|]-1}_{i=1,2,3,\dots} g^{s t}_{i}(m,n;\lambda) V^{s+t-i}_{m+n}.
\end{equation}
With the constants:
\begin{equation}
\label{eq:strcons}
g^{s t}_{i}(m,n;\lambda)\equiv \frac{q^{i-2}}{2(i-1)!} 
\,_4F_3 \left[
\begin{array}{llll} 
\frac{1}{2}+\lambda & \quad \frac{1}{2}-\lambda \quad \frac{2-i}{2}\quad \frac{1-i}{2}\\
\frac{3}{2}-s & \quad \frac{3}{2}-t \quad \frac{1}{2}+s+t-i
\end{array} \bigg|1\right] N^{st}_{i}(m,n),\end{equation}
$q=\frac{1}{4}$ and : 
\begin{equation}
\label{eq:Nstru}
N^{st}_{i}(m,n)=\scriptstyle\sum_{k=0}^{i-1}(-1)^{k} \begin{pmatrix}i-1\\ k\end{pmatrix} \bigl(s-1+m+1\bigr)_{k-i+1} \bigl(s-1-m+1\bigr)_{-k} \bigl(t-1+n+1\bigr)_{-k} \bigl(t-1-n+1\bigr)_{k-i+1}.
\end{equation}
The  $(n)_{k}$ are the ascending  Pochhammer symbols. 
We define trace as: 
\begin{equation}
\tr \left(V^s_{m_s}V^s_{-m_s}\right)\equiv\frac{6}{1-\lambda^2}
\frac{(-1)^{m_s} 2^{3-2s}\Gamma(s+m_s)\Gamma(s-m_s)}{(2s-1)!!(2s-3)!!}
\prod^{s-1}_{\sigma=1}\left(\lambda^2-\sigma^2\right). \label{trace}
\end{equation}
\section{Uniqueness of the choice $\nu_0=\frac{1}{2}$, $\nu_{i>0}=0$ for $0<\lambda<1$}\label{sec:M}
\noindent
Here we show how the only solution to the integrability condition (\ref{quantis}) in the region $0<\lambda<1$ is the trivial one $n_0=1$. First we write down the first $4\times4$ block of the upper triangular matrix $M$ 
\begin{equation}
\left(
\begin{array}{cccc}\small
 1 & \frac{4 (\lambda^2-4) }{15} & \frac{4(\lambda^2-4) \left(11 \lambda ^2-71\right)}{315}  & \frac{4 
  (\lambda^2-4) \left(107 \lambda ^4-1630 \lambda ^2+6563\right)}{4725} \\
 0 & \frac{12 \prod^{3}_{\sigma=2}   \sqrt{(\lambda^2-\sigma^2)}}{5 \sqrt{14}} & \frac{4 \left(7 \lambda ^2-67\right) \prod^{3}_{\sigma=2}   \sqrt{(\lambda^2-\sigma^2)}}{15 \sqrt{14}} & \frac{4
\prod^{3}_{\sigma=2}   \sqrt{(\lambda^2-\sigma^2)} \left(893 \lambda ^4-19090 \lambda ^2+113957\right)}{2475 \sqrt{14}} \\
 0 & 0 & \frac{8 \sqrt{\frac{5}{11}}\prod^{5}_{\sigma=2}   \sqrt{(\lambda^2-\sigma^2)}}{21} & \frac{80\sqrt{\frac{5}{11}}\prod^{5}_{\sigma=2}   \sqrt{(\lambda^2-\sigma^2)} \left(5 \lambda ^2-89\right) }{819} \\
 0 & 0 & 0 & \frac{32 \sqrt{\frac{7}{5}}\prod^{7}_{\sigma=2}   \sqrt{(\lambda^2-\sigma^2)}}{429} 
\end{array}
\right) .\label{M}
\end{equation}
The eigenvalues can be checked to be greater or equal than one in $0<\lambda<1$. In fact they grow as the diagonal index $i$ grows. Next we show this excludes the presence of any other solution. Be the following definition and couple of facts
\begin{equation}
{n_O}^i\equiv O^i_j n^j,  ~ ~  O M^T M O^T= Diag((M^{i  i})^2), ~~ O^T O= 1.
\end{equation}
As $(M^{ii})^2\geq1$ it is clear that 
\begin{equation}
\sum^\infty_{i=1}\left( \left(M \centerdot n \right)^i\right)^2=\sum^\infty_{i=1}\left( M^{ii}\right)^2 \left({n_{O}}_i\right)^2 \geq \sum^{\infty}_{i=1} {n_O}_i^2= \sum^{\infty}_{i=1} {n}_i^2\geq 1. \label{s}
\end{equation}
The saturation in (\ref{s}) comes when one of the integers $n_i$ is $\pm 1$. As $(M^{ii})^2=1$ only if $i=1$ thence the only solution to (\ref{quantis}) is the trivial one. Notice however that our conclusions do breakdown when we are out of the region
$0<\lambda<1$. This is, to define a new solution we just need to tune up $\lambda$ in such a way that for a given $i$, $M^{ii}=\pm1$.

 \section{Solutions with dynamical critical exponent $z<1$} \label{sec:zm1}
Here we study the fluctuations for a specific background with dynamical critical exponent $z<1$. We take as a toy example the case $\bar{\mu}_3=-\mu_3\neq0$. The secular polynomial reads out
 \begin{eqnarray}
  i k=i k^\prime_r-2\mu_3\left(\omega^2+k^{\prime 2}_r+\frac{\lambda^2-1}{3}\right),\label{eqkz<1}
  \end{eqnarray}
whose roots are
\begin{equation}
k^\prime_{\pm}=\frac{-i+\sqrt{-1+8 i k \mu _3-\frac{16}{3} \left(\lambda ^2+3 \omega ^2-1\right) \mu _3^2}}{4 \mu _3}.\label{kz<1}
\end{equation}
From the quantisation condition \eqref{quantisationCond}
\begin{eqnarray}\label{QNMlif1}
w^{\pm}_{1-n}=-i \frac{1}{2}\left(1+2n +\lambda\right)+{\delta^{\pm}_1}_{z<1},\nonumber\\
w^{\pm}_{2-n}=-i \frac{1}{2}\left(1+2n +\lambda\right)+{\delta^\pm_{2}}_{z<1},
\end{eqnarray}
where the $\pm$ refer to the $\pm$ in \eqref{kz<1} and the $(1,2)$ refer to the $(+,-)$ in \eqref{quantisationCond} respectively, and
\begin{eqnarray}\small
{\delta^{\pm}_1}_{z<1}&=\frac{3 i\mp \sqrt{-1+8 (-1+2 i k-2 n-\lambda ) \mu _3+\frac{16}{3} \left(5+12 n^2+6 \lambda +\lambda ^2+12 n (1+\lambda )\right) \mu _3^2}}{8 \mu _3}, \nonumber\\
   {\delta^{\pm}_2}_{z<1}&=\frac{-3 i\pm \sqrt{-1+8 (1+2 i k+2 n+\lambda ) \mu _3+\frac{16}{3} \left(5+12 n^2+6 \lambda +\lambda ^2+12 n (1+\lambda )\right) \mu _3^2}}{8 \mu _3}.
\end{eqnarray} \\
We can also study the case $\bar{\mu}_3=\mu_3$, we get in this case from \eqref{secular}:
\begin{equation}
k'=\frac{k+4 i k \omega  \mu _3}{1+16 \omega ^2 \mu _3^2}.
\label{eq:kpmu=-mub}
\end{equation}
We get just one root, which means that after the folding process of section \ref{sec:ex}, the final equation obtained is of second order, as can be explicitly checked. The quasi normal modes in this case are given by:
\begin{eqnarray}
\omega_{1\pm}&=\frac{-i-4 i (1+2 n+\lambda ) \mu _3\mp\sqrt{-1+8 (1-2 i k+2 n+\lambda ) \mu _3-16 (1+2 n+\lambda )^2 \mu _3^2}}{8 \mu _3},\nonumber\\
\omega_{2\pm}&=\frac{-i-4 i (1+2 n+\lambda ) \mu _3\mp\sqrt{-1+8 (1+2 i k+2 n+\lambda ) \mu _3-16 (1+2 n+\lambda )^2 \mu _3^2}}{8 \mu _3}.\label{QNMlif2}
\end{eqnarray}

In  section \ref{sec:BH} we have given the metric  for these solutions \eqref{eq:lifini}. Propagation in Lifshitz metrics with $z<1$ is typically associated with the presence of superluminal excitations in the dual field theory, see for instance \citep{Hoyos:2010at, Koroteev:2009xd}. For each one of our blocks $r$ we can make use of the AdS/CFT dictionary. The dispersion relations for the corresponding physical excitation, $n$, is given by the condition for a pole in the retarded 2-point function (\ref{quantisationCond}) and the expression for the auxiliary momentum $k^\prime_r$ of the given block in terms of $k$ and $w$ are given in \eqref{kz<1} and (\ref{eq:kpmu=-mub}) respectively.  The wavefront velocity $v_f
=\lim_{\omega\rightarrow \infty}\frac{\omega}{k_R(\omega,n)}$,  \citep{Amado:2008ji}, can be computed to be
\begin{equation}\label{eq:KQNMc-a}
v_{f1}=\lim_{\omega\rightarrow \infty}\frac{\omega}{-\omega +4 \omega  \mu _3+8 n \omega  \mu _3+4 \lambda  \omega  \mu _3}=\frac{1}{-1 +4 \mu _3+8 n \mu _3+4 \lambda \mu _3 },
 \end{equation}
\begin{equation}\label{eq:KQNMc-b}
v_{f2}=\lim_{\omega\rightarrow \infty}\frac{\omega}{\omega +4 \omega  \mu _3+8 n \omega  \mu _3+4 \lambda  \omega  \mu _3} =\frac{1}{1 +4 \mu _3+8 n \mu _3+4 \lambda \mu _3}.
 \end{equation}
We end up by noticing that for $ |\mu_3|\geq\frac{1}{2(1+\lambda)}$ there are no superluminal modes ($|v_f|\leq1$) in these examples. But for other values there is a finite number of them. However the tale of large $n$ excitations have all $|v_f|\leq1$.  \\ \\

\section{Differential operators and $C_{BTZ}$  }\label{sec:HOEqs}

We present some differential operators that were referenced in the main body of the text. The Klein Gordon operator in $\rho$ coordinates:
\begin{eqnarray}\small
D_2\equiv\frac{d^2}{d\rho^2}&+\frac{2 (e^{4 \rho }+1)}{(e^{4 \rho }-1)} \frac{d}{d\rho}+\frac{(1-\lambda ^2)(e^{8\rho}-1)}{(e^{4 \rho }-1)^2} -\frac{2\left(2(k^2-\omega^2)( e^{2 \rho} +e^{6 \rho })+\lambda^2-1-e^{4 \rho } (4 k^2+4 \omega^2+\lambda^2-1)\right)}{(e^{4 \rho }-1)^2}.\nonumber\\
\label{D2rho}
\end{eqnarray}
The operator $D_4$ for the background $\mu_3\neq 0$
\begin{eqnarray}
D_4(z)&\equiv \partial_z^4 -\frac{2 i w (z-1)+2 (\lambda -4) z+4}{(z-1) z}\partial_z^3+\left(\frac{-3 (z-1) z+6 i \mu_3 (z-1) z (k+2 w)}{12 \mu_3^2 (z-1)^2 z^2}\right. ~~~~~~~~~~~~~~~~~~~~~~~~~~~~\nonumber\\& \left.-\frac{3 w^2 (z-1)^2-9 i w (z-1) ((\lambda -3) z+1)+z ((\lambda -18) \lambda -(\lambda -4) (4 \lambda -11) z+44)-6}{3 (z-1)^2 z^2}\right)\partial_z^2~~~~~~~~~\nonumber\\&+\frac{(w (z-1)-i ((\lambda -2) z+1)) (6 k \mu_3+4 \mu_3 (3 w+(\lambda -2) \mu_3 (3 w-i (\lambda -4)))+3 i)}{12 \mu_3^2 (z-1)^2
   z^2} \partial_z ~~~~~~\nonumber\\&-\frac{(-i (\lambda -1) (2 (\lambda -2) \mu_3+3)+3 k+3 w) \left(-i (\lambda -1) (2 (\lambda -2) \mu_3-3)+3 k+12 i \mu_3 w^2+3 w (4
   (\lambda -1) \mu_3-1)\right)}{144 \mu_3^2 (z-1)^2 z^2}.~~~~~~~  \label{D4}
\end{eqnarray}
 The differential operator $\overset{(1)}{D}_{GK}$ that we make reference to in section \eqref{sec:ChGau} 
\begin{eqnarray}
&&D^{(1)}_{GK}=\frac{64 i e^{2 \rho }(3 e^{2 \rho }-1) k}{(e^{2 \rho }-1)^2 (1+e^{2 \rho })^3 (\lambda ^2-1)}\frac{d}{d\rho}\nonumber \\ &+&\frac{8 k \left(\frac{1-11 k^2-\omega^2-\lambda ^2+ e^{6 \rho }(-7k^2+3 \omega^2-5\lambda ^2-11)}{(e^{2 \rho }-1)^3} +e^{4\rho } (3 k^2+9 \omega^2+\lambda ^2-1)\right)}{-i e^{-2 \rho } (1+e^{2 \rho })^4 (\lambda ^2-1)} \nonumber \\ &+& \frac{8k\left(\frac{e^{8 \rho }(42\omega^2+6k^2+2\lambda^2-2)+ e^{4 \rho }(29-15k^2+59 \omega^2+3\lambda ^2)+e^{2 \rho } (27 k^2+25 \omega^2+\lambda ^2-17)}{(e^{2 \rho }-1)^3}\right)}{-i e^{-2 \rho }(1+e^{2 \rho })^4 (\lambda ^2-1)}.\label{eq:D1KGP}
\end{eqnarray}

Finally, we give the master field  $C$ for the BTZ$_{M=-1}$ background up to spin 4. We have used the Fourier basis \eqref{eq:four} and redefined $ C^1_0\equiv C$:
\begin{eqnarray}
 C^{2}_{\pm1}&=&\frac{6 i e^{\rho } \left(\mp(e^{2 \rho }-1) k+(e^{2
       \rho }+1) \omega\right) C[\rho ]}{(e^{2 \rho }-1)(e^{2 \rho
   }+1)(\lambda ^2-1)},\label{C21}\\
C^{2}_0&=& -\frac{6C'[\rho ]}{\lambda ^2-1}, \label{C20}\\
C^{3}_0&=& \frac{30\left(\frac{6(k^2-\omega^2)( e^{2 \rho }+ e^{6 \rho })}{\lambda ^2-1}
    +1+e^{8 \rho }-2 e^{4 \rho } (\frac{6 k^2+6
    \omega^2}{\lambda ^2-1}+1)\right)C[\rho]}{(e^{4 \rho}-1)^2 (\lambda ^2-4)} \nonumber \\
&-&\frac{90(e^{8 \rho }-1) C'[\rho]}{(e^{4 \rho}-1)^2 (4-5 \lambda ^2+\lambda ^4)},\label{C30} 
\end{eqnarray}
\begin{eqnarray}
C^{3}_{\pm1}&=& \frac{\left(\frac{\mp(e^{3 \rho }- e^{\rho })}{(1+e^{2 \rho })^2}
      k+ \omega\frac{(e^{3 \rho }+ e^{\rho })}{(e^{2 \rho }-1)^2}\right)C[\rho ]+
    \left(\frac{\pm  e^{\rho }}{(1+e^{2 \rho })} k-\frac{ e^{\rho
        }}{(e^{2 \rho }-1)} \omega \right)C'[\rho ]}{\frac{(4-5
    \lambda ^2+\lambda ^4)}{60 i}}, \label{C31}\\
C^{3}_{\pm2}&=& -\frac{30 \left(\frac{\mp e^{\rho }}{(e^{2 \rho }+1)} k+\frac{ e^{\rho }}{(e^{2 \rho }-1)}\omega\right)^2 C[\rho ]+\frac{30 e^{2 \rho }}{(e^{4 \rho }-1)}C'[\rho ]}{(4-5 \lambda ^2+\lambda ^4)} \label{C32}\\
C^{4}_{0}&=& \frac{\left(( e^{2 \rho }+4
      e^{6 \rho }+ e^{10 \rho })\frac{(k^2-\omega^2)}{\lambda^2-1}+\left(\frac{1+e^{12 \rho
        }}{8}-(e^{4 \rho }+e^{8\rho }) (\frac{3 k^2+3 \omega^2}{\lambda
          ^2-1}+\frac{1}{8})\right)\right)C[\rho]}{\frac{(e^{4 \rho
    }-1)^3(\lambda ^2-9)(\lambda ^2-4)}{5600}}\nonumber \\
&-& \frac{  \left(( e^{2 \rho }+ e^{6 \rho
    })(k^2-\omega^2)+\frac{(1+e^{8 \rho })(11+\lambda ^2)}{10}-2 e^{4
      \rho }( k^2+ \omega^2+\frac{\lambda ^2-29}{10}\right)C'[\rho ]}{
  \frac{(e^{4 \rho }-1)^2(\lambda ^2-9)(\lambda ^2-4)(\lambda
    ^2-1)}{42000}},\nonumber\\\label{C40}\end{eqnarray}
    \begin{eqnarray}
C^{4}_{\pm1}&=&\left(\frac{\pm k\left(\frac{(1+\lambda ^2)(1+e^{8 \rho })}{5}- (e^{2 \rho }+ e^{6 \rho })(2+\omega^2)-2
        e^{4 \rho }(\omega^2+\frac{\lambda ^2-9}{5})\right)}{\frac{ i e^{-\rho }(e^{2
        \rho }-1)^2(e^{2 \rho }+1)^3 (\lambda ^2-9)(\lambda ^2-4)(\lambda
    ^2-1)}{2100}}\right. \nonumber\\ &+&\left.\frac{ \pm e^{2 \rho }k^3- e^{2 \rho }\frac{ (e^{2 \rho }+1)}{(e^{2
        \rho }-1)}
      k^2 \omega-\frac{(e^{2 \rho }+1)^3}{(e^{2
        \rho }-1)^3} \omega \left(\frac{(1+\lambda ^2)(1+e^{4 \rho })}{5}+\frac{e^{2 \rho
        } (8-5 \omega^2-2 \lambda ^2)}{5}\right)}{\frac{i e^{-\rho} (e^{2 \rho }+1)^3 (\lambda ^2-9)(\lambda ^2-4)(\lambda ^2-1)}{2100}}\right) C[\rho]\nonumber\\ &-& \frac{2(e^{2 \rho }-1) \left(\pm(e^{2 \rho }-e^{4 \rho
      }+\frac{e^{6\rho }-1}{2}) k-( e^{2 \rho }+ e^{4 \rho }+\frac{e^{6 \rho }+1}{2})
      \omega \right)C'[\rho]}{\frac{i e^{-\rho}(e^{2 \rho }+1)^2 (\lambda ^2-9)(\lambda ^2-4)(\lambda ^2-1)}{2100}},\label{C41}\end{eqnarray}
      
      \begin{eqnarray}
      C^4_{\pm2}&=&-420e^{2\rho}\left(\frac{\pm8k \omega+(1-\lambda^2\mp4k\omega+ 4\omega^2)(1+e^{8 \rho })+2 e^{4 \rho } (1+20\omega^2)}{(e^{4 \rho }-1)^3 (\lambda ^2-9) (\lambda ^2-4)(\lambda ^2-1)}\right. \nonumber \\ &+& \left. \frac{\left(20 e^{4 \rho }-12( e^{2 \rho } +e^{6 \rho })+2(1+e^{8 \rho })\right)(k^2-\omega^2)}{(e^{4 \rho }-1)^3 (\lambda ^2-9) (\lambda ^2-4)(\lambda ^2-1)}\right) C[\rho ] \nonumber \\ &+&420e^{2\rho}\frac{ \left(\pm 4k \omega-2 e^{2 \rho } (k^2-\omega^2)+(1+e^{4 \rho }) (k^2\mp2k \omega+\omega^2-4)\right) C'[\rho ]}{(e^{4 \rho }-1)^2 (\lambda ^2-9) (\lambda ^2-4) (\lambda ^2-1)},\nonumber\\\label{C42}
\end{eqnarray}
\begin{eqnarray}
C^4_{\pm3}&=&\frac{\left(\frac{\pm k \left(3 \omega^2+e^{4 \rho }(3 \omega^2-2)+e^{2 \rho } (4+6 \omega^2)-2\right)}{(e^{2 \rho }-1)^2}\pm k^3-\frac{3(1+e^{2 \rho }) k^2 \omega}{e^{2 \rho }-1}-\frac{(1+e^{2 \rho })^3 \omega (\omega^2-2)}{(e^{2 \rho }-1)^3}\right)C[\rho ]}{\frac{ -i e^{-3 \rho }  (e^{2 \rho }+1)^3 (\lambda ^2-9)(\lambda ^2-4) (\lambda ^2-1)}{140}}\nonumber\\&+&\frac{\left(\pm(e^{2 \rho}-1) k-(1+e^{2 \rho }) \omega\right)C'[\rho ]}{\frac{-i e^{3 \rho }(e^{4 \rho }-1)^3 (\lambda ^2-9)(\lambda ^2-4) (\lambda ^2-1)}{420}}.\label{C43}
\end{eqnarray}

 The primes stand for derivative along $\rho$, and one  can  recover the result in coordinate space ($t,\phi$) by replacing $k\rightarrow-i\partial_\phi$ and $\omega\rightarrow-i\partial_t$. Notice that all these higher spin components are generically singular
at the horizon. \\

\bibliographystyle{JHEP}
\bibliography{HSlambdaQNM}
\end{document}